\documentclass[journal]{IEEEtran}
\usepackage{amsmath}
	\usepackage{amsmath, amsthm, amssymb}
	\usepackage{graphicx}
	\usepackage{epstopdf}
	\usepackage{algpseudocode}%
	\usepackage{float}
	\usepackage{graphicx}
	\usepackage{amsmath}
	\usepackage{latexsym}
	\usepackage{graphicx}
	\usepackage{bm}
	\usepackage{amssymb}
	\usepackage{array} 
	\usepackage{setspace}
	\usepackage{fancyhdr}
	\usepackage{epstopdf}
 \input epsf
	\usepackage{xcolor}
	\usepackage{multicol}
	\usepackage{lipsum}
	\usepackage{multicol,lipsum}
	\usepackage{caption}
	\usepackage{verbatim}
	\usepackage{calligra}
  \usepackage{placeins}
	\usepackage[english]{babel} 
\usepackage [autostyle, english = american]{csquotes}
\usepackage{amsmath,amsfonts,amsthm,bm} 
	\usepackage[font={footnotesize}]{caption}
	\usepackage[scr=boondoxo,
	scrscaled=1.15,]{mathalfa}
	\usepackage{multirow}
	\usepackage{algorithm}
\usepackage{algpseudocode}
\usepackage{algpascal}
\usepackage{tikz}
\usepackage{pgfplots}
\usepackage{graphicx}
\pgfplotsset{compat=newest}
\usetikzlibrary{shapes,arrows,fit,positioning,calc}
\usetikzlibrary{plotmarks}
\usetikzlibrary{decorations.pathreplacing}
\usetikzlibrary{patterns}
\usetikzlibrary{chains,arrows,shapes,spy}
\usetikzlibrary{automata}
\usepackage{algpseudocode}

\def\BibTeX{{\rm B\kern-.05em{\sc i\kern-.025em b}\kern-.08em
T\kern-.1667em\lower.7ex\hbox{E}\kern-.125emX}}
\usepackage{balance}

\begin{document}
\alglanguage{pseudocode}

\title{Decentralized and Centralized IDD Schemes for Cell-Free Networks\\
}
\pdfinclusioncopyfonts=1
		\author{\IEEEauthorblockN{
				{Tonny~Ssettumba},~
		{Zhichao~Shao},~  {Lukas~T.~N.~Landau,} and {Rodrigo~C.~de~Lamare} \vspace{-0.25em}
				}		\\	\thanks{T. Ssettumba, L. T. N. Landau and R. C. de Lamare are with the Center for Telecommunications Studies (CETUC), Pontifical Catholic University of Rio de Janeiro (PUC-Rio). E-mails:  tssettumba@gmail.com, \{landau, delamare\}@puc-rio.br.
     Z.~Shao is
with the National Key Laboratory of Science and Technology on Communications, University of Electronic Science and Technology of
China, Chengdu 611731, China (e-mail: zhichao.shao@uestc.edu.cn).} 
    }
\maketitle
\begin{abstract}
In this paper, we propose iterative interference cancellation schemes with access points selection (APs-Sel) for cell-free massive multiple-input multiple-output (CF-mMIMO) systems. Closed-form expressions for centralized and decentralized linear minimum mean square error (LMMSE) receive filters with APs-Sel are derived assuming imperfect channel state information (CSI). Furthermore, we develop a list-based detector based on LMMSE receive filters that exploits interference cancellation and the constellation points. A message-passing-based iterative detection and decoding (IDD) scheme that employs low-density parity-check (LDPC) codes is then developed. Moreover, log-likelihood ratio (LLR) refinement strategies based on censoring and a linear combination of local LLRs are proposed to improve the network performance.  
We compare the cases with centralized and decentralized processing in terms of bit error rate (BER) performance, complexity, and signaling under perfect CSI (PCSI) and imperfect CSI (ICSI)  and verify the superiority of the distributed architecture with LLR refinements. 
\end{abstract}
\section{Introduction}
\IEEEPARstart{C}{ell}-free massive multiple-input multiple-output (CF-mMIMO) networks exploit the distributed nature of large-scale multiple-antenna systems to improve the quality of service, yielding very high throughput \cite{rr1}. In such networks, user equipments (UEs) are served by a large number of access points (APs), which are equipped with either single antenna or multiple antennas \cite{rr2, r9,itermmsecf,rmmsecf}. Also, the distributed nature of the network with extra spatial degrees of freedom makes the channel between UEs and APs almost orthogonal, which reduces the level of interference 

\subsection{Prior and Related Works}

In the traditional  CF-mMIMO systems all APs serve  UEs  in the entire service area \cite{rr1,rr2,rr3,rr4}. However,  such a setting is highly impractical and non-scalable since it requires increased front haul link connections between the APs and the central processing unit (CPU). Moreover, the hardware complexity of the network grows exponentially with the increase in radio frequency units and signaling \cite{rr5,spa,rr61xx,rr62xx,cesg,rscf,clust&sched,RR13,mmimo,wence}. Recently,  access points selection (APs-Sel) techniques  have  been proposed to reduce signaling and the number of front haul connections , whose performance is close to that of the traditional CF-mMIMO system \cite{rr1,rr4,rr5}. 
For example, the work in \cite{rr4} proposed a sort and connect algorithm  between 
based on the effective channel gain (ECG) and channel quality. A joint power allocation algorithm is proposed to minimize the total energy consumption and reduce overhead signaling for APs selection \cite{rr5}. This approach is compared with other APs-Sel algorithms based on the largest large-scale fading (LLSF) coefficients \cite{rr2}. 

Moreover, another factor that influences the performance of CF-mMIMO systems is the  multi-user interference (MUI) due to their broadcast nature, pilot contamination as well as overlapping of the uplink transmitted signals \cite{rr2,mmimo,wence}. 
To address this issue, an efficient receiver design is necessary \cite{rr6,rrv6, r1,r2,r3,r3v,spa,spa2,spa1,spa3,spa4,spa5,spa_r,mbdf+o,r4,r5,spa,itic,mfsic,dfcc,mbdf,did,lrcc}.   Prior works on receiver designs in CF-mMIMO \cite{rr1,rr2,rr6,rrv6} have studied linear receivers such as receive matched filter (RMF) and minimum mean square error (MMSE) for the uncoded system. However, the MUI remains constant  even at high SNR regime. To further reduce the MUI, channel codes with iterative detection and decoding (IDD) schemes are presented in \cite{r1,r2,r3,spa,r4,r5,RR10,RR11,RR12,RR13,r10,r11,r14,spa,jed,mfsic,mbdf,aaidd,listmtc,detmtc,msgamp,msgamp2,dynovs}.  In which channel codes  with IDD can correct   unreliable detection decisions as the number of iterations increases. Moreover, using long code word  yields improved performance at the expense of increased complexity in terms of time and computing power \cite{RR11,RR12}.   IDD techniques use the message-passing principle by exchanging the soft beliefs in  form of log-likelihood ratios (LLRs) between the detector and the decoder \cite{r1, r2, r5,mbdf,r7,r8,llraps,refidd} . 

In IDD schemes, soft interference cancellation (Soft-IC) detectors are generally used due to their low complexities. For example, the list-based detection techniques that are capable of eliminating error propagation are presented in \cite{r5,mbdf}.  
In \cite{r7}, a local partial marginalization detector based on turbo iterations was proposed for uplink CF-mMIMO systems.
The proposed detector was compared with other baseline schemes, such as MMSE  and MMSE with SIC (MMSE-SIC) detectors. 
The authors in \cite{r77} proposed a distributed expectation detector for CF-mMIMO systems, where the CPU is equipped with a non-linear module and the APs are equipped with a linear module. 
Before sending the posterior mean and variance to the CPU, the APs first detect the symbols using local CSI. The extrinsic data for each AP is then generated and integrated at the CPU using maximum-ratio combining (MRC).  Shaik \emph{et al.} \cite{RR11x} performed a theoretical analysis on the distributed  computation of LLRs based on the optimal maximum a posterior (MAP) detector for sequential architecture of  CF-mMIMO. However, the complexity of such an optimal filter grows exponentially with the increase in the number of UEs and antennas, which is prevalent in CF-mMIMO networks. Moreover, the 
 bit error rate (BER) performance of the proposed network was not presented in terms of numerical results.
 \subsection{Motivation and Contributions}
 Based on literature, a few works are devoted 
 to the BER performance analysis for coded CF-mMIMO networks using message passing  for various implementation architectures and cooperation levels \cite{r10,r11, RR13}. There is still a need for low-complexity LLR processing schemes for decentralized processing in CF-mMIMO networks to facilitate BER performance close to the centralized schemes. Also, list-based detection methods like MF-SIC have not been investigated for CF-mMIMO systems, despite the fact that they have the potential to improve performance by eliminating the error propagation that occurs in traditional MMSE-SIC schemes \cite{llraps,refidd}. Additionally, the use of MMSE receivers with soft-IC (MMSE- Soft-IC) detectors and LPDC codes 
 can produce very simple, efficient and practical implementation 
 compared to optimal detectors. 
Therefore, this  work studies iterative centralized  and decentralized CF-mMIMO architectures, which is different from the work in \cite {r7} that considers non-iterative processing. We also propose three LLR processing strategies for low complexity detection schemes. The main contributions of this paper can be summarized as follows:

\begin{itemize}

\item IDD schemes with LDPC channel codes for scalable and non-scalable centralized and decentralized CF-mMIMO schemes are developed.

\item New closed-form expressions for the MMSE-Soft-IC detectors are derived for the case with APs-Sel while taking the channel estimation errors into account. Based on the derived equations,  general expressions are deducted for a system that uses all the APs (All-APs). Furthermore,  we draw insights and derive the uplink MMSE receive filters based on the a-priori information of transmitted bits. 

\item A novel List-MMSE-Soft-IC detector to reduce the error propagation in the IC step is proposed. The  List-MMSE-Soft-IC performance is compared with other baseline detectors such as LMMSE-Soft-IC and soft LMMSE. 

\item Three novel  LLR processing schemes are proposed for decentralized CF-mMIMO:  a) one  based on decoding the LLRs at each AP (Standard LLR processing); b)  another based on censoring the LLRs by decoding each user equipment (UE) information at the AP, where it achieves the largest mean absolute value of LLRs (LLR Censoring). This censoring of LLRs helps to reduce redundant processing at the CPU and c) the last that provides refinements in the LLRs by performing a linear combination of LLRs from the different APs  by assuming statistical independence of the APs. 

\item The proposed local LLR processing strategies are compared with the traditional network, which is based on centralized processing schemes.  The impact of different IDD iterations is also examined. Furthermore, the performance of the decentralized and centralized architectures is compared in terms of computational complexity and the amount of required signaling between the APs and CPU.
\end{itemize}

\subsection{ Paper Organization and Notation}

The rest of this paper is organized as follows:
Section \ref{sys} presents the proposed centralized and decentralized IDD processing schemes, receiver designs, channel estimation statistics, as well as the list-based detector.  The LLR statistics, refinement, computational complexity, signaling loads, and the considered decoding algorithm are presented in Section \ref{IDD}. Numerical results, network setup, assumptions, and remarks are presented in Section \ref{SIM_RESULTS}. Finally, Section \ref{Concl} gives the conclusions. 

\textit{Notation:} Lower and upper bold case symbols  represent vectors and matrices, respectively. The operator $(\cdot)^{H}$ denotes the Hermitian transpose.

\section{Proposed System Model}\label{sys}
We consider the uplink of a CF-mMIMO system with $L$ APs and $K$ single-antenna user equipment (UEs), where each AP is equipped with $N$ receive antennas. In this section, we describe  two proposed IDD-based systems, which are centralized IDD scheme described in Subsection \ref{CentralizedDetector} and decentralized IDD scheme described in  Subsection \ref{decentralized}.  
 
 \subsection{Proposed Centralized IDD Scheme}\label{CentralizedDetector}
 
 In this section, we present the proposed 
centralized processing architecture for coded CF-mMIMO. The code word sequence $\bm{c}_{k}$ is created by first encoding the message sequence $\bm{m}_{k}$  for UE $k$ by an LDPC encoder (Enc) with a code rate of $R$. This encoded sequence is then modulated (Mod) to form  complex symbols with complex constellation of $2^{M_{c}}$ possible signal points. The ${K}$ UEs then send the  modulated symbols to the APs.
 The APs serve as relays during data reception and transfer the information to the CPU, which has a joint detector (Det), LLR computing module and an LDPC decoder (Dec). Then, the  detector forwards the data to a module that computes the LLRs $\Lambda_{i}$. These computed LLRs are then sent to the decoder. By providing extrinsic data $\Lambda_{e}$ to the joint detector, the decoder uses an iterative technique presented in Section \ref{IDD} that enhances the detection performance of the receiver.
 \subsubsection{Uplink Pilot Transmission and Channel Estimation}\label{channel_est}

We assume that $\tau_{p}$ mutually orthogonal pilot sequences $\bm{\psi}_{1}$, ..., $\bm{\psi}_{\tau_{p}}$ with $||\bm{\psi}_{t}||^{2} = \tau_{p}$ are used to estimate the channel. Furthermore, 
 $K>\tau_{p}$ is such that more than one UE can be assigned per pilot. The index of UE $k$ that uses the same pilot is denoted as $t_{k} \in \left\{1,..., \tau_{p}\right\}$ with $\vartheta_{k} \subset\left\{1, ..., K\right\}$ as the subset of UEs that use the same pilot as UE $k$ inclusive. 
The complex received signal at the $l$-th AP after the UE transmission, \cite{rr1,rr2, r11} $\mathbf{Y}_{l}$, with dimensions  $N\times\tau_{p}$, is given by
 \begin{align}\label{chanesti}
\mathbf{Y}_{l}=\sum_{j=1}^{K}\sqrt{\eta_{j}}\mathbf{g}_{jl}\bm{\psi}^{T}_{t_{j}}+\mathbf{N}_{l},
 \end{align}
 where $\eta_{j}$ is the transmit power from UE $j$, $\mathbf{N}_{l}$ is the received noise signal with independent ${\mathcal{N}}_{\mathbb{C}}\sim\left(0, \sigma^{2}\right)$ entries and noise power $\sigma^{2}$, $\mathbf{g}_{jl}\sim\mathcal{N}_{\mathbb{C}}\left(0, \bm{\Omega}_{jl}\right)$, and $\bm{\Omega}_{jl}\in\mathbb{C}^{N\times N}$ is the spatial correlation matrix that describes the channel's spatial properties between the $k$-th UE and the $l$-th AP, $\beta_{k,l}\triangleq \frac{\mathrm{tr}\left(\bm{\Omega}_{jl}\right)}{N}$ is the large-scale (LS) fading coefficient. The first AP 
 correlates the received signal with the associated normalized pilot signal $\bm{\psi}_{t_{k}}/\sqrt{\tau_{p}}$ to 
$\mathbf{y}_{t_{kl}}\triangleq \frac{1}{\sqrt{\tau_{p}}}\mathbf{Y}_{l}\bm{\psi}^{*}_{t_{k}}\in \mathbb{C}^{N}$ to estimate the channel $\mathbf{g}_{jl}$ given by
\begin{align}\label{eq2}
  \mathbf{y}_{t_{kl}}=\sum_{j\in \vartheta_{k}}\sqrt{\eta_{j}\tau_{p}}\mathbf{g}_{jl}+\mathbf{n}_{t_{kl}},  
\end{align}
where $\mathbf{n}_{t_{kl}}\triangleq \frac{1}{\sqrt{\tau_{p}}}\mathbf{N}_{l}\bm{\psi}^{*}_{t_{k}}\sim\mathcal{N}_{c}\left(0, \sigma^{2}\mathbf{I}_{N}\right)$ is the obtained noise sample after estimation. 
From \cite{rr1}, the MMSE estimate of $\mathbf{g}_{kl}$ is given by 
\begin{align}  \hat{\mathbf{g}}_{kl}=\sqrt{\eta_{k}\tau_{p}}\bm{\Omega}_{kl}\Psi^{-1}_{t_{kl}}\mathbf{y}_{t_{kl}},
\end{align}
where $\Psi_{t_{kl}}=\mathbb{E}\left \{\mathbf{y}_{t_{kl}}\mathbf{y}^{H}_{t_{kl}} \right \}=\sum_{j\in \vartheta_{k}}{\eta_{j}\tau_{p}}\mathbf{\Omega}_{jl}+\mathbf{I}_{N}$ is the received signal vector correlation matrix. The channel estimate $\hat{\mathbf{g}}_{kl}$ and the estimation error $\tilde{\mathbf{g}}_{kl}=\mathbf{g}_{kl}-\hat{\mathbf{g}}_{kl}$ are independent with distributions $\hat{\mathbf{g}}_{kl}\sim\mathcal{N}_{c}\left(0, \eta_{k}\tau_{p}\bm{\Omega}_{kl}\mathbf{\Psi}^{-1}\bm{\Omega}_{kl}\right)$ and $\tilde{\mathbf{g}}_{kl}\sim\mathcal{N}_{c}\left(0, \mathbf{C}_{kl}\right)$, where the matrix $\mathbf{C}_{kl}$ is given by
\begin{align}
   \mathbf{C}_{kl}=\mathbb{E}\left \{ \tilde{\mathbf{g}}_{kl} \tilde{\mathbf{g}}^{H}_{kl}\right \}=\bm{\Omega}_{kl}-\eta_{k}\tau_{p}\bm{\Omega}_{kl}\mathbf{\Psi}^{-1}\bm{\Omega}_{kl}.
\end{align}
It should be noted that the pilot contamination is caused by the mutual interference made by UEs using the same pilot signals in \eqref{eq2}, which lowers the system's performance \cite{rr1}.

From \eqref{chanesti}, the received signal vector after stacking  the channel vectors from all the APs is given by
\begin{align}\label{receiv_CPU_signal}   \mathbf{y}=\mathbf{G}\mathbf{s}+\mathbf{n},
 \end{align}
where the channel matrix $\mathbf{G}\in \mathbb{C}^{NL\times K}$ has both small scale and LS fading coefficients. Vector $\mathbf{s}=\left[s_{1},..,s_{K}\right]^{T}  $ is the transmitted symbols with  $\mathbb{E}\left\{s_{k}s^{*}_{k}\right\}=\rho_{k}$, the average transmit power is given by  $\bm{\rho}=\left  [ \rho_{1},..,  \rho_{K} \right ]  ^  {T}$,   $\mathbf{n}$ is the additive white Gaussian noise (AWGN).

In CF-mMIMO networks, there are limitations on the complexity and amount of signaling the APs and CPU must exchange. Both of these issues make system modeling and design almost impracticable. To solve this problem, we adopt a scalable CF-mMIMO setup that takes the selection of APs into account. This is accomplished 
using the APs-selection technique described as follows.
\subsubsection{Access Point Selection Procedure}
The dynamic cooperative clustering (DCC) approach described in \cite{rr2, r11} is 
considered when selecting the APs.  By letting $\mathcal {M}_{k}\subset \left\{1,...,L\right\}$ be the subset of APs in service of UE $k$, the matrix $\mathbf{D}_{kl}$ is defined as
\begin{align}
\mathbf{D}_{kl}=\left\{\begin{matrix}
\mathbf{I}_{N} & \text{if} & l\in \mathcal {M}_{k}\\ 
 \mathbf{0}_{N}& \text{if}  & l\not\in \mathcal {M}_{k}.
\end{matrix}\right.
\end{align}
 The APs that provide service to a specific UE are determined by the  block diagonal matrix $\mathbf{D}_{k}=\mathrm{diag}\left(\mathbf{D}_{k1},..,\mathbf{D}_{kL}\right)\in \mathbb{C}^{NL\times NL}$.  
When $\mathbf{D}_{k}=\mathbf{I}_{NL}$, it  implies that all APs 
 serve all the UEs. However, using all APs is not scalable and practical and thus clustering approaches such as user-centric techniques can be adopted.
Then, the set of UEs that are served by AP $l$ is 
\begin{align}
    \mathcal{D}_{l}=\biggl\{k:\mathrm{tr}\left(\mathbf{D}_{kl}\right)\geq 1,k\in\left\{1,..,K\right\}\biggr\}.
\end{align} 
It is important to note that the DCC does not alter the  received signal statistics since all APs receive the   broadcast signal. An essential feature of such a selection process is limiting the number of APs that take part in signal detection.
The joint APs selection criterion described in \cite{rr2} 
determines which APs will provide service to a specific UE. In this scenario, the UE designates a master AP to coordinate uplink (UL) detection and decoding based on the LLSF. The CPU then establishes a threshold value $\beta_{th}$ for non-master APs to provide service to a certain UE. A detailed explanation 
of the operation of the DCC approach can be found in \cite{rr2}. 

There is a need to demodulate the transmitted symbols at the receiver. Thus, the proposed centralized  receive filters and structure derivation are given as follows. 

\subsubsection{Proposed Centralized Receiver Design}
The proposed receiver configuration aims to cancel the MUI 
caused by the other $K-1$  UEs in the network. The  receiver consists of an MMSE filter followed by a soft interference cancellation scheme, which may use either a successive  or a list-based successive-interference cancellation technique. The receiver first creates soft estimates of the transmitted symbols by computing the symbol mean $\bar{s}_{j}$ based on the soft beliefs from the LDPC decoder \cite{r3,r10,r11} given $\mathbb{E}\left \{ s_{j} \right \}= \bar{s_{j}}$ as described by
\begin{align}\label{expectation_sym}
	   \bar{s}_{j}=\sum_{s\in\mathcal{A}}s P(s_{j}=s),
\end{align}
where  $\mathcal{A}$  is the set of complex constellations.
The a-priori probability of the extrinsic LLRs is given by
\begin{align}\label{aprior_prob}
	 P(s_{j}=s)=\prod_{l=1}^{M_{c}}\lbrack 1+\exp(-s^{b_{l}}\Lambda_{i}(b_{(j-1)M_{c}+l}))\rbrack^{-1},
\end{align}
where $\Lambda_{i}(b_{i})$ is the extrinsic LLR of the $i$-th  bit calculated by the LDPC decoder from the previous iteration, and $s^{b_{l}}\in (+1,-1)$ denotes   the $l$-th  bit of symbol $s$. 
The variance of the $j$-th UE symbol is calculated as
	\begin{align}\label{variex}
	\sigma^{2}_{j} =\sum_{s\in \mathcal{A}}|s-\bar{s}_{j}|^{2}P(s_{j}=s). 
	\end{align} 
The received signal after decomposition of \eqref{receiv_CPU_signal} is given by
 \begin{align}\label{receiv_CPU_signalAPSEL}
\mathbf{y}=\hat{\mathbf{g}}_{k}s_{k}+\hat{\mathbf{G}}_{\text{i}}\mathbf{s}_{\text{i}}+\sum_{m=1}^{K}\tilde{\mathbf{g}}_{m}s_{m}+\mathbf{n},
 \end{align}
 where the first term on the right-hand side (RHS)  
 is the desired signal, the second term is the interference from the other $K-1$ users, the third term denotes the interference from the channel estimation and the fourth 
 is the phase-rotated noise.
 After the estimated MUI has been removed and APs have been selected, the received symbol estimate of the  $k$-th UE data stream at the CPU is given by
\begin{align}\label{funAPS1CENT}  \tilde{s}_{k}=\mathbf{w}^{H}_{k}{\mathbf{D}}_{k}\left( \mathbf{y}-\hat{\mathbf{G}}_{\text{i}}\bar{\mathbf{s}}_{\text{i}}\right),
\end{align}
The optimization of the receive combining filter $\mathbf{w}_{k}$ is achieved by minimizing the mean square error (MSE) between the symbol estimate and the transmitted symbol. The formulation of the optimization problem is given by
\begin{align}\label{funp2lp}
   \mathbf{w}_{k}   =\mathsf{arg}\min_{\left ( \mathbf{w}_{k}\right )}\mathbb{E}\biggl\{||\tilde{s}_{k}-s_{k}||^{2}\mid \hat{\mathbf{G}}\biggr\}.
\end{align}
Differentiating the objective function on the RHS of \eqref{funp2lp} with respect to (w.r.t) $\mathbf{w}^{H}_{k}$, the optimal MMSE receive filter  $\mathbf{w}_{k}$ should satisfy the following relation
\begin{align}\label{objxx}
{\mathbf{D}}_{k} \mathbb{E}\{\mathbf{y}_{R}\mathbf{y}^{H}_{R}\}{\mathbf{D}}^{H}_{k}\mathbf{w}_{k}-{\mathbf{D}}_{k}\mathbb{E}\{\mathbf{y}_{R}{s}^{*}_{k}\}=0,
\end{align}
where $\mathbf{y}_{R}=\mathbf{y}-\hat{\mathbf{G}}_{\text{i}}\bar{\mathbf{s}}_{\text{i}}$. The solution to the filter is obtained by making $\mathbf{w}_{k}$ the subject of \eqref{objxx} as
\begin{align}\label{fcentC}
    \mathbf{w}_{k} =\left({\mathbf{D}}_{k} \mathbb{E}\{\mathbf{y}_{R}\mathbf{y}^{H}_{R}\}{\mathbf{D}}^{H}_{k}\right)^{-1}{\mathbf{D}}_{k}\mathbb{E}\{\mathbf{y}_{R}{s}^{*}_{k}\}.
\end{align}
By using the orthogonality principle in \cite{RR14} and assuming statistical independence of each term in the received signal $\mathbf{y}$, the terms $ \mathbb{E}\{\mathbf{y}_{R}{s}^{*}_{k}\}$ and $  \mathbb{E}\{\mathbf{y}_{R}\mathbf{y}^{H}_{R}\}$ are given by 
\begin{align}\label{w1C}
  \mathbb{E}\{\mathbf{y}_{R}{s}^{*}_{k}\}=\rho_{k}\hat{\mathbf{g}}_{k},
  \end{align}
\begin{align}\label{w2C}
    &\mathbb{E}\{\mathbf{y}_{R}\mathbf{y}^{H}_{R}\}=\rho_{k}\hat{\mathbf{g}}_{k}\hat{\mathbf{g}}^{H}_{k}+\hat{\mathbf{G}}_{\text{i}}\Delta_{\text{i}}\hat{\mathbf{G}}^{H}_{\text{i}}\notag\\&+\sum_{m=1}^{K}\left ( \mid s_{m}\mid^{2} +\sigma_{m}^{2}\right )\mathbf{C}_{m}+\sigma^{2}\mathbf{I}_{NL}.
\end{align}
The matrix $\mathbf{\Delta}_{\text{i}}=\mathsf{diag}\left[{\sigma_{{1}}^{2}},...,{\sigma_{{k-1}}^{2}},  {\sigma_{{k+1}}^{2}},...,{\sigma_{{K}}^{2}}\right]$ denotes the covariance matrix that consists of the entries computed in \eqref{variex}. 
By substituting \eqref{w1C} and \eqref{w2C} into \eqref{fcentC}, the centralized MMSE filter is given by
\begin{align}\label{det_2}
  &\mathbf{w}_{k}= \rho_{k}\biggl ({\mathbf{D}}_{k}\biggl(\rho_{k}\hat{\mathbf{g}}_{k}\hat{\mathbf{g}}^{H}_{k}+\hat{\mathbf{G}}_\text{i}\mathbf{\Delta}_\text{i}\hat{\mathbf{G}}^{H}_\text{i}\biggr){\mathbf{D}}^{H}_{k}\notag\\&+{\mathbf{D}}_{k}\biggl(\sigma^{2}\mathbf{{I}}_{NL}+\sum_{m=1}^{K}\left(\mid{s}_{m}\mid^2+\sigma^{2}_{m}\right)\mathbf{C}_{m}\biggr){\mathbf{D}}^{H}_{k}\biggr )^{-1}{\mathbf{D}}_{k}\hat{\mathbf{g}}_{k}.
\end{align}
A detailed derivation of \eqref{det_2} can be found in Appendix A.
\subsubsection{Insights into the Derived Centralized MMSE Filter}
For scenarios involving the usage of all APs, PCSI, and different number of IDD iterations, we draw insights into the derived MMSE receive filter given in \eqref{det_2}.

For the case with all APs, the selection matrix $\mathbf{D}_{k}=\mathbf{I}_{NL}$. Thus, the filter in \eqref{det_2} is given by 
\begin{align}\label{cent_filter}
\mathbf{w}_{k} =&\biggl(\rho_{k}\hat{\mathbf{g}}_{k}\hat{\mathbf{g}}^{H}_{k}+\hat{\mathbf{G}}_{\text{i}}\Delta_{\text{i}}\hat{\mathbf{G}}^{H}_{\text{i}}+\sum_{m=1}^{K}\left ( \mid s_{m}\mid^{2} +\sigma_{m}^{2}\right)\mathbf{C}_{m}\notag\\&+\sigma^{2}\mathbf{I}_{NL}\biggr)^{-1}\rho_{k}\hat{\mathbf{g}}_{k}.
\end{align}
For the case with PCSI, the channel estimation error $\tilde{\mathbf{g}}_{m}$ is $0$, which makes $\mathbf{C}_{m}=0$. This yields $\hat{\mathbf{g}}_{k}=\mathbf{g}_{k}$.
Thus, the third term in \eqref{cent_filter} vanishes to zero, and we obtain  
 \begin{align}\label{eqlpobjfun122}
\mathbf{w}_{k}&=\rho_{k}\biggl ( \rho_{k}\mathbf{g}_{k}\mathbf{g}^{H}_{k}+\mathbf{G}_{\text{i}}\Delta_{\text{i}}\mathbf{G}^{H}_{\text{i}}+\sigma^{2}\mathbf{I}_{NL} \biggr )^{-1}\mathbf{g}_{k}.
 \end{align}
In the first iteration it is considered that $\bar{\mathbf{s}}_{\text{i}}=\mathbf{0}$ in \eqref{expectation_sym}. In this case, we have a linear MMSE filter and the estimated signal in \eqref{funAPS1CENT} is 
\begin{align}\label{MMSE_DP}
&\tilde{s}_{k}=\rho_{k}\hat{\mathbf{g}}^{H}_{k}{\mathbf{D}}^{H}_{k}\biggl(\rho_{k}{\mathbf{D}}_{k}\hat{\mathbf{g}}_{k}\hat{\mathbf{g}}^{H}_{k}{\mathbf{D}}^{H}_{k}+{\mathbf{D}}_{k}\hat{\mathbf{G}}_\text{i}~\mathrm{diag}\left(\bm{\rho}_\text{i}\right)\hat{\mathbf{G}}^{H}_\text{i}{\mathbf{D}}^{H}_{k}\notag\\&+{\mathbf{D}}_{k}\left(\sigma^{2}\mathbf{I}_{NL}+\sum_{m=1}^{K}\rho_{m}\mathbf{C}_{m}\right){\mathbf{D}}^{H}_{k}\biggr )^{-1}\mathbf{y}.
\end{align}
The vector $\bm{\rho}_{\text{i}}$ denotes the average transmit power vector for the other $K-1$ UEs. 

As the number of iterations increases, there is more a posterior information about the transmitted bit. This implies that mean of the interference symbol  $\bar{\mathbf{s}}_{\text{i}}\approx\mathbf{s}_{\text{i}}$ in \eqref{expectation_sym}.  Thus,  the filter becomes a perfect interference canceler, and  \eqref{funAPS1CENT} yields 
 \begin{align}\label{iterinc}
&\tilde{s}_{k}=\rho_{k}\hat{\mathbf{g}}^{H}_{k}{\mathbf{D}}^{H}_{k}\biggl (\rho_{k} {\mathbf{D}}_{k}\hat{\mathbf{g}}_{k}\hat{\mathbf{g}}^{H}_{k}{\mathbf{D}}^{H}_{k}\notag\\&+{\mathbf{D}}_{k}\biggl(\sigma^{2}\mathbf{I}_{NL}+\sum_{m=1}^{K}\mid s_{m}\mid^{2}\mathbf{C}_{m}\biggr){\mathbf{D}}^{H}_{k} \biggr )^{-1}\left ( \mathbf{y}-{\mathbf{D}}_{k}\hat{\mathbf{G}}_\text{i}\mathbf{s}_\text{i} \right ).
 \end{align}
The centralized detection schemes experience high levels of complexity as the number of UEs, APs and antennas at the APs increase. This makes the design of receivers more complicated. Also the amount of required signaling between APs and the CPU increases. The proposed decentralized filter and its structure are given as follows. 
\subsection{Proposed Decentralized IDD Scheme }\label{decentralized}
¨


The proposed decentralized IDD scheme works as follows. The transmitter operates in the same way as that of the centralized processor in Subsection \ref{CentralizedDetector}. The key difference is explained as follows.
The coded data blocks are  transmitted by ${K}$ UEs to the APs. Each AP is equipped with a local detector, LLR computing module and an LDPC decoder. The APs use their local channel estimates to perform IDD on the received signals. The detector sends its symbol estimates to a module that computes the local soft information $\lambda_{i}$ in the form of LLRs. The computed LLRs are then sent to the decoder, which performs iterative processing by exchanging extrinsic information $\lambda_{e}$ with the local detector. 
In the decentralized operation, the APs act as compute-and-forward relays by sending their soft beliefs to the CPU for further processing. The challenge at the CPU is to design 
an intelligent way of processing these LLRs. We devise three techniques for processing these LLRs. The first scheme (Standard LLR processing) is based individual decisions from each AP and an average BER is then computed based on these decisions from each AP. The second scheme considers  censoring the LLRs (LLR Censoring) and decoding each UE data at the AP where it achieves the largest mean absolute value of LLRs. The third scheme scheme is based on the linear combination of the LLRs (LLR Ref). A detailed explanation, operation and analysis  of the proposed LLR processing schemes is given in Section \ref{IDD}. 
 
\subsubsection{Proposed Decentralized Receiver Design}\label{RCA}
The channel statistics, estimation, and received signals follow the model introduced in Subsection \ref{channel_est}. 
The received signal at the $l$-th AP is given by
\begin{align}\label{rx_lp}   \mathbf{y}_{l}=\sum_{i=1}^{K}\mathbf{g}_{il}s_{i}+\mathbf{n}_{l}\in\mathbb{C}^{N\times 1},
\end{align}
which  can further be decomposed as
\begin{align}\label{decomposed_sign}  \mathbf{y}_{l}=\hat{\mathbf{g}}_{kl}s_{k}+\hat{\mathbf{G}}_{\text{i}l}\mathbf{s}_{\text{i}}+\sum_{m=1}^{K}\tilde{\mathbf{g}}_{ml}s_{m}+\mathbf{n}_{l},
\end{align}
where the first term on the RHS 
is the desired signal, the second term is the interference from the other $K-1$ users, the third term denotes the interference due to channel estimation errors and the fourth term 
is the phase-rotated noise. 
The received local symbol estimate of the $k$-th UE data stream at the $l$-th AP after removing the MUI is given by
\begin{align}\label{funAPS1} \tilde{s}_{kl}=\mathbf{w}^{H}_{kl}{\mathbf{D}}_{kl}\left( \mathbf{y}_{l}-\hat{\mathbf{G}}_{\text{i}l}\bar{\mathbf{s}}_{\text{i}}\right),
\end{align}
where the notation ${\mathbf{D}}_{kl}$ implies that the $l$-th AP is among the selected APs.  Here, the optimization of the received combining filter $\mathbf{w}_{kl}$ is obtained by minimizing the error between the estimated detected symbol and the transmitted symbol. The optimization problem is  formulated as
\begin{align}\label{funp2lp11}
   \mathbf{w}_{kl}   =\mathsf{arg}\min_{\left ( \mathbf{w}_{kl}\right )}\mathbb{E}\biggl\{||\tilde{s}_{kl}-s_{k}||^{2}\mid \hat{\mathbf{G}}_{l}\biggr\}.
\end{align}
The derivation is similar to the centralized approach and, for completeness, it is described in detail in Appendix B. The optimal receive filter $\mathbf{w}_{kl}$ should satisfy the following relation
\begin{align} {\mathbf{D}}_{kl}\mathbb{E}\{\mathbf{y}_{Rl}\mathbf{y}^{H}_{Rl}\}{\mathbf{D}}^{H}_{kl}\mathbf{w}_{kl}-{\mathbf{D}}_{kl}\mathbb{E}\{\mathbf{y}_{Rl}{s}^{*}_{k}\}=0.
\end{align}
Thus, the solution to the receive filter is given by
\begin{align}\label{fdec}
    \mathbf{w}_{kl} =\left({\mathbf{D}}_{kl}\mathbb{E}\{\mathbf{y}_{Rl}\mathbf{y}^{H}_{Rl}\}{\mathbf{D}}^{H}_{kl}\right)^{-1}{\mathbf{D}}_{kl}\mathbb{E}\{\mathbf{y}_{Rl}{s}^{*}_{k}\}.
\end{align}
The terms $ \mathbb{E}\{\mathbf{y}_{Rl}{s}^{*}_{k}\}$ and $  \mathbb{E}\{\mathbf{y}_{Rl}\mathbf{y}^{H}_{Rl}\}$ are respectively, given by 
\begin{align}\label{w1d}
  \mathbb{E}\{\mathbf{y}_{Rl}{s}^{*}_{k}\}=\rho_{k}\hat{\mathbf{g}}_{kl},
  \end{align}
\begin{align}\label{w2d}
    &\mathbb{E}\{\mathbf{y}_{Rl}\mathbf{y}^{H}_{Rl}\}= \rho_{k}\hat{\mathbf{g}}_{kl}\hat{\mathbf{g}}^{H}_{kl}+\hat{\mathbf{G}}_{\text{i}l}\Delta_{\text{i}l}\hat{\mathbf{G}}^{H}_{\text{i}l}\notag\\&+\sum_{m=1}^{K}\left ( \mid s_{m}\mid^{2} +\sigma_{m}^{2}\right )\mathbf{C}_{ml}+\sigma^{2}\mathbf{I}_{N} , 
\end{align}
where 	 $\Delta_{\text{i}l}$ denotes the covariance matrix that consists of the entries computed in \eqref{variex}, locally computed at the $l$-th AP.
 By substituting \eqref{w1d} and \eqref{w2d} into \eqref{fdec}, the solution of the local receive filter $\mathbf{w}^{*}_{kl}$ is given by
\begin{align}\label{eqlpobjfun1}
&\mathbf{w}_{kl}=\rho_{k}\biggl ( {\mathbf{D}}_{kl}\biggl(\rho_{k}\hat{\mathbf{g}}_{kl}\hat{\mathbf{g}}^{H}_{kl}+\hat{\mathbf{G}}_{\text{i}l}\Delta_{\text{i}l}\hat{\mathbf{G}}^{H}_{\text{i}l}+\sigma^{2}\mathbf{I}_{N}\notag\\&+\sum_{m=1}^{K}\left ( \mid s_{m}\mid^{2} +\sigma_{m}^{2}\right )\mathbf{C}_{ml}\biggr){\mathbf{D}}^{H}_{kl} \biggr )^{-1}{\mathbf{D}}_{kl}\hat{\mathbf{g}}_{kl}.
 \end{align}
 A detailed derivation of this filter can be found in Appendix B.
 \subsubsection{Insights into the Derived Decentralized MMSE Filter}
In what follows, we draw insights into the derived local MMSE filter expression for the cases with a selected AP, PCSI and different number of iterations.
When the local AP is selected with all its 
antennas, the selection matrix $\mathbf{D}_{kl}=\mathbf{I}_{N}$. The local  filter in \eqref{eqlpobjfun1} is 
\begin{align}\label{dcentallaps}
&\mathbf{w}_{kl}=\rho_{k}\biggl ( \rho_{k}\hat{\mathbf{g}}_{kl}\hat{\mathbf{g}}^{H}_{kl}+\hat{\mathbf{G}}_{\text{i}l}\Delta_{\text{i}l}\hat{\mathbf{G}}^{H}_{\text{i}l}\notag\\&+\left(\sum_{m=1}^{K}\left ( \mid s_{m}\mid^{2} +\sigma_{m}^{2}\right )\mathbf{C}_{ml}+\sigma^{2}\mathbf{I}_{N}\right) \biggr )^{-1}\hat{\mathbf{g}}_{kl}.
 \end{align}
Assuming PCSI, the fourth term in \eqref{eqlpobjfun1} vanishes to zero and we obtain the filter given by 
 \begin{align}\label{eqlpobjfun122}
\mathbf{w}_{kl}&=\rho_{k}\biggl ( \rho_{k}{\mathbf{D}}_{kl}\mathbf{g}_{kl}\mathbf{g}^{H}_{kl}{\mathbf{D}}^{H}_{kl}+{\mathbf{D}}_{kl}\mathbf{G}_{\text{i}l}\Delta_{\text{i}l}\mathbf{G}^{H}_{\text{i}l}{\mathbf{D}}^{H}_{kl}\notag\\&+\sigma^{2}{\mathbf{D}}_{kl}\mathbf{I}_{N} \tilde{\mathbf{D}}^{H}_{kl}\biggr )^{-1}\mathbf{g}_{kl}{\mathbf{D}}_{kl}.
 \end{align}
In the first iteration, 
$\bar{\mathbf{s}}_{\text{i}}=\mathbf{0}$ in \eqref{expectation_sym}, which yields the linear MMSE filter and the estimated signal in \eqref{funAPS1} given by 
\begin{align}\label{MMSE_DP}
&\tilde{s}_{kl}=\rho_{k}\hat{g}^{H}_{kl}{\mathbf{D}}^{H}_{kl}\biggl({\mathbf{D}}_{kl}\biggl(\rho_{k}\hat{\mathbf{g}}_{kl}\hat{\mathbf{g}}^{H}_{kl}+ \hat{\mathbf{G}}_{\text{i}l}\mathrm{diag}\left({\rho}\right)\hat{\mathbf{G}}^{H}_{\text{i}l}\notag\\&+  \sum_{m=1}^{K} \left ( \mid\bar{s}_{m}\mid^{2}+\sigma^{2}_{m} \right )\mathbf{C}_{ml}+  \sigma^{2}\mathbf{I}_{N}\biggr){\mathbf{D}}^{H}_{kl}\biggr )^{-1}\mathbf{y}_{l},
\end{align}
where the parameter $\bm{\rho}$ denotes the average transmit power vector for the other $K-1$ UEs. 
The mean symbol 
$\bar{\mathbf{s}}_{\text{i}}\approx\mathbf{s}_{\text{i}}$ in \eqref{expectation_sym} as the number of iterations increases. Thus  the filter becomes a perfect interference canceler,  \eqref{funAPS1} yields 
 \begin{align}\label{iterinc}
\tilde{s}_{kl}=&\rho_{k}\hat{\mathbf{g}}^{H}_{kl}{\mathbf{D}}^{H}_{kl}\biggl ( {\mathbf{D}}_{kl}\biggl( \rho_{k}\hat{\mathbf{g}}_{kl}\hat{\mathbf{g}}^{H}_{kl}+  \sum_{m=1}^{K} \mid{s}_{m}\mid^{2}\mathbf{C}_{ml}\notag\\&+  \sigma^{2}\mathbf{I}_{N}\biggr){\mathbf{D}}^{H}_{kl}\biggr )^{-1}\left ( \mathbf{y}_{l}-{\mathbf{D}}_{kl}\hat{\mathbf{G}}_{il}\mathbf{s}_\text{i} \right ).
 \end{align}
The derived MMSE-SIC filters in \eqref{det_2} and \eqref{eqlpobjfun1} experience error propagation at each sequential step. In what follows, we describe the proposed list-based detector that is capable of suppressing the  error that occurs at each cancellation step.
 \subsection{Proposed List-based detector} \label{MFSIC}
In this subsection, we detail the proposed list-based detection scheme that is 
similar to the one adopted in our previous papers \cite{r10,r11}. 


The design takes advantage of list feedback (LF) diversity by selecting a list of constellation candidates if there is the unreliability of the previously detected symbols \cite{r10,r11,mfsic,dfcc}. 
A shadow area constraint (SAC) is employed 
to obtain an optimal feedback candidate. The SAC is capable of reducing the search space from exponentially growing 
as well as reducing 
computational complexity. The key idea of such a selection criterion is to avoid redundant processing when there is a reliable decision.
The procedure of obtaining the detected symbol $\hat{s}_{k}$ of the $k$-th user is analogous to the steps presented in \cite{r10,r11,r5}. The $k$-th user soft estimate is obtained by $u_{k}=\mathbf{w}_{k}^{H}{\mathbf{D}}_{k}\check{\mathbf{y}}_{k}$. The filter $\mathbf{w}_{k}$ is the receive MMSE filter described in \eqref{fcentC} 
 and later \eqref{fdec}. The residual signal $\check{\mathbf{y}}_{k}=\mathbf{y}-\sum_{t=1}^{k-1}\hat{\mathbf{g}}_{t}\hat{{s}}_{t}$ is the received vector following the soft cancellation of the $k-1$ symbols that were previously detected. The SAC assesses the reliability of this decision using the soft estimate $u_{k}$ for each layer according to
\begin{align}
    d_{k}=\vert u_{k}-\nu_{f}\vert,
\end{align}
where $\nu_{f}=\mathsf{arg}\min_{\mathbf{\nu_{f}}\in{\mathcal{A}}}  \left\{ \vert u_{k}-\nu_{f}\vert\right\}$ denotes the closest constellation point to the $k$-th user soft estimate $u_{k}$. If $d_{k}>d_{\text{th}}$, the chosen constellation point gets dumped into the shadow area of the constellation map since the choice is deemed to be unreliable. The parameter $d_{\text{th}}$ is the predefined threshold on the Euclidean distance to ensure the reliability of the selected symbol \cite{r10, r11}. The list-based algorithm performs hard slicing for UE $k$ as in the soft-IC if 
the soft estimate $u_{k}$ is reliable. In this case, $\hat{s}_{k}=Q(u_{k})$ is the estimated symbol, where $Q(\cdot)$ is the quantization notation which maps to the constellation symbol closest to  $u_{k}$.

Otherwise, the decision is deemed unreliable and a candidate list  $\mathcal{L}=\{c_{1}, c_{2},...,c_{m},...,c_{M}\}\subseteq\mathcal{A}$ is generated, which is made up of the $M$  constellation points that are closest to $u_{ k}$, where $M\le 2^{{M}_{c}}$. 
The algorithm selects an optimal candidate $c_{m,\text{opt}}$ from a list of $\mathcal{L}$ candidates. Thus, the unreliable choice $Q(u_{k})$ is replaced by a hard decision and $\hat{s}_{k}=c_{m,\text{opt}}$ is obtained. 
The list-based detector first defines the selection vectors $\bm{\phi}^{1},\bm{\phi}^{2},...,\bm{\phi}^{m},...\bm{\phi}^{M}$ whose size is equal to the number of the constellation candidates that are used every time a decision is considered unreliable. For example, for the $k$-th  layer,  a $K\times 1$ vector $\bm{\phi}^{m}
   =\left [ \hat{s}_{1},...,\hat{s}_{k-1}, c_{m},\phi^{m}_{k+1},...,\phi^{m}_{q},...,\phi^{m}_{K} \right ]^T$ which is  a potential choice corresponding to $c_{m}$ in the $k$-th user comprising the following items: (a) the previously estimated symbols $\hat{s}_{1},\hat{s}_{2}, ..., \hat{s}_{k-1}$; (b) the candidate symbol $c_{m}$ obtained from the constellation for subtracting a decision that was considered unreliable $Q(u_{k})$ of the $k$-th user; (c) using (a) and (b) as the previous decisions, detection of the next user data $k+1, ...,q,...,$ $K$-th is performed by the soft-IC approach. Mathematically, the choice $\phi^{m}$ is given by \cite{r11}
\begin{align}
 \phi^{m}_{q}=Q(\mathbf{w}^{H}_{q}{\hat{\bm{y}}}^{m}_{q}),
\end{align}
where the index $q$ denotes a UE between the $k+1$-th and the $K$-th UE, $    \hat{\bm{y}}^{m}_{q}=\check{\bm{y}}_{k}-{\mathbf{D}}_{k}\hat{\mathbf{g}}_{k}c_{m}- {\mathbf{D}}_{k}\sum_{p=k+1}^{q-1}\hat{\mathbf{g}}_{p}\phi^{m}_{p}$. A key attribute of the list-based detector is that the same MMSE filter $\mathbf{w}_{k}$ is used for all the constellation candidates. Therefore, it has a computational cost that is close to that of the conventional soft-IC.
The optimal candidate 
$m_{\text{opt}} $ is selected according to the local maximum likelihood (ML) rule given by 
\begin{align} {m_{\text{opt}}}=\mathsf{arg}\min_{1\leq m\leq M}\left \|{\mathbf{D}}_{k}\bm{y}-{\mathbf{D}}_{k}\hat{\mathbf{G}}\bm{\phi}^{m}\right\|^{2}_{2}.
\end{align}
The derived centralized and decentralized filters suffer from interference due to the other $K-1$ users, channel estimation errors and AWGN noise. This makes the derived filters not to be Gaussian because the output of a Gaussian filter should be Gaussian, for a Gaussian input. In the next section, we approximate the filters to be Gaussian  by computing the mean and variances, present the  LLR processing schemes, perform signaling and computational complexity analysis as well explaining the considered decoding algorithm.
\section{Iterative Detection and Decoding}\label{IDD}
This section presents the IDD schemes for the studied  MMSE-based detectors,  consisting of a detector and an LDPC decoder. The received signal at the output of the  receive filter contains the desired symbol, MUI, and noise. The parameter $u_{k}$ is assumed to be an output of an AWGN channel \cite{r7, r10} given by
\begin{align}\label{idd1}
  u_{k}=\omega_{k}s_{k}+z_{k},
\end{align}
where 
$\mathbb{E}\{{s}^{*}_{k}u_{k}\}=\rho_{k}\mathbf{w}^{H}_{k}\tilde{\mathbf{D}}_{k}\hat{\mathbf{g}}_{k}$ and $\mathbb{E}\{{s}^{*}_{k}u_{kl}\}=\rho_{k}\mathbf{w}^{H}_{kl}\tilde{\mathbf{D}}_{kl}\hat{\mathbf{g}}_{kl}$ are for the centralized and decentralized schemes, respectively. Iterative techniques to calculate the MMSE receive filter are also possible \cite{jidf,jiols,jiomimo}.The parameter $z_{k}$ is a zero-mean AWGN variable.
Using similar procedures as in \cite{r7,r10}, the variance $\kappa^{2}$ of $z_{k}$ is computed by $\kappa^{2}=\mathbb{E}\left\{\mid u_{k}-\omega_{k}s_{k}\mid^{2}\right\}$:
\begin{align}\label{sdemapcent}
\kappa^{2}=\mathbf{w}^{H}_{k}{\mathbf{D}}_{k}\left(\hat{\mathbf{G}}_{\text{i}}\Delta_{\text{i}}\hat{\mathbf{G}}^{H}_{\text{i}}+\sum_{m=1}^{K}\rho_{m}\mathbf{C}_{m}+\sigma^{2}\mathbf{I}_{NL} \right ){\mathbf{D}}^{H}_{k}\mathbf{w}_{k}, 
\end{align} and 
\begin{align}\label{sdemapdeccent}
\kappa^{2}_{l}=\mathbf{w}^{H}_{kl}{\mathbf{D}}_{kl}\left ( \hat{\mathbf{G}}_{\text{i}l}\Delta_{\text{i}}\hat{\mathbf{G}}^{H}_{\text{i}l}+\sum_{m=1}^{K}\rho_{m}\mathbf{C}_{ml}+\sigma^{2}\mathbf{I}_{N} \right ){\mathbf{D}}^{H}_{kl}\mathbf{w}_{kl},
\end{align}
are for the centralized and decentralized schemes, respectively. 

Detailed derivations of \eqref{sdemapcent} and \eqref{sdemapdeccent} are presented in Appendix C.
The extrinsic LLR computed by the detector for the $l$-th bit $l\in\left\{1,2,...,M_{c}\right\}$ of the symbol $s_{k}$  \cite{r10,r11} is given by
\begin{align}\label{LLR_COMP}
   &\Lambda_{e}\left ( b_{(k-1)M_{c}+l} \right )=\frac{\log P\left ( b_{(k-1)M_{c}+l}=+1 |u_{k}\right)}{\log P\left ( b_{(k-1)M_{c}+l}=-1| u_{k}\right)}\notag\\&-\frac{\log P\left ( b_{(k-1)M_{c+1}}=+1 \right )}{\log P\left ( b_{(k-1)M_{c+1}}=-1 \right )} \notag\\&
   =\log\frac{\sum _{s\in A^{+1}_{l}}f\left ( u_{k}|s \right )P\left (s \right )}{\sum _{s\in A^{-1}_{l}}f\left ( u_{k}|s \right )P\left (s \right )}-\Lambda_{i}\left ( b_{(k-1)M_{c}+l} \right ),
\end{align}
where the last equality of \eqref{LLR_COMP} follows from Bayesian rule. The parameter  $A^{+1}_{l}$ is the set of $2^{Mc-1}$ hypothesis  for which the $l$-th bit is $+1$.   l  The a-priori probability $P(s)$ is given by \eqref{aprior_prob}. The approximation of the likelihood function \cite{r10,r11} $f(u_{k}|s)$ is given by
    \begin{align}\label{llfn}
        f\left ( u_{k}|s \right )\simeq\frac{1}{\pi\kappa^{2}}\exp\left (-\frac{1}{\kappa^{2}} |u_{k}-\omega_{k}s|^{2} \right ).
    \end{align}
   After local processing, the CPU has to perform final decisions by using the LLRs from the different APs. This is accomplished by proposing three LLR processing strategies presented as follows.
     \subsection{Standard LLR Processing}
In this strategy, each AP computes the BER based on decisions from its LLRS. After obtaining the BER from each AP, an average BER is calculated for the entire network.  However, such an approach yields poor results as some APs have very unreliable LLRs for particular UEs. We then discuss  two proposed strategies to improve the performance of local detectors.

\subsection{LLR Censoring}
In this subsection we present an LLR censoring technique that helps to reduce the redundant processing of LLRs at the CPU. 
First, the independent streams of LLRs are sent from the APs to the CPU with dimensions $KC_{\text{leng}}$. At each AP, we compute the mean absolute value of the LLRs which is given by
\begin{align}
\mu_{\Lambda_{kl,e}}=\frac{1}{C_\text{leng}}\sum_{c=1}^{C_{\text{leng}}}\rvert\Lambda_{l,e}\rvert.
\end{align}
Based on $\mu_{\Lambda_{kl,e}}$, the UE is decoded at the AP when this parameter is highest and the other LLRs are discarded.  This is done for all APs and a new matrix $\bm{\Lambda^{\text{new}}_{k,e}}$ with the censored LLRs is formed and used in performing final decoding. The  LLR censoring strategy is summarized in Algorithm \ref{alg}.
\begin{algorithm}
\caption{Algorithm for Censoring Local LLRs}\label{alg}
\begin{algorithmic}
\State $\bm{\Lambda}_{e}\in\mathbb{C}^{KC_{\text{leng}}L}$, $\bm{\Lambda^{\text{new}}_{k,e}}=\bm{0}_{KC_{\text{leng}}}$
\For{l=1 to L}
\For{k=1 to K}
\If{$\mu_{\Lambda_{kl,e}}\geq\max\left(\mu_{\bm{\Lambda}_{k,e}}\right)$}
\State$\bm{\Lambda^{\text{new}}_{k,e}}=\bm{\Lambda}_{kl,e}$
\Else \State Continue
\EndIf
\State  $k\gets k+1$
\EndFor
\State  $l\gets l+1$
\EndFor
\State Output $\bm{\Lambda^{\text{new}}_{k,e}}$
\end{algorithmic}
\end{algorithm}

\subsection{LLR Refinement}

We propose an LLR refinement strategy that computes the linear summation of the multiple streams of LLRs obtained from the locally computed joint IDD detectors. Mathematically, the refined combination of LLRs at the CPU is given by
\begin{align}\label{LLR_AVG}
\Lambda_{\rm avg,e} \left( b_{(k-1)M_{c}+l} \right) = 
\sum_{l=1}^{L} \Lambda_{l,e} \left( b_{(k-1)M_{c}+l} \right).
\end{align}
The idea of combining multiple streams of LLRs creates some diversity benefits from the LLRs  
and yields some performance improvement in the network. Another key advantage 
of decentralized processing is that each AP has accurate channel estimates; 
thus, it is better to perform the detection locally than at the CPU. The mean of the refined LLRs is given by
\begin{align}\label{LLR_mean}
\mathbb{E}[\Lambda_{\rm avg,e} \left( b_{(k-1)M_{c}+l} \right)] & =  \sum_{l=1}^{L} \mathbb{E}[\Lambda_{l,e} \left( b_{(k-1)M_{c}+l} \right)],\\
& = \mu_{\Lambda_{\rm avg,e} },
\end{align}
where $\mathbb{E}[\Lambda_{l,e} \left( b_{(k-1)M_{c}+l} \right)] \rightarrow 0$ since 
\begin{equation}
  \begin{split}
\mathbb{E}[\Lambda_{l,e}\left( b_{(k-1)M_{c}+l} \right)]&=\log \int  \Lambda_{l,e}\left(b_{(k-1)M_{c}+l}\right)\notag\\&\times p_{\Lambda_{l,e}|H_0} d \Lambda_{l,e}, \\ & = \log \int  \frac{p_{\Lambda_{l,e}|H_1}}{p_{\Lambda_{l,e}|H_0}}  p_{\Lambda_{l,e}|H_0} d \Lambda_{l,e} = 0, \nonumber
\end{split}
\end{equation} 
where $p_{\Lambda_{l,e}|H_1}$ is the conditional probability density function (pdf) of the LLR of stream $l$ given bit $1$ and $p_{\Lambda_{l,e}|H_0}$ is the conditional pdf of the LLR given bit $0$.

The variance of the refined LLRs is given by
\begin{equation}
\begin{split}
\sigma^2_{\Lambda_{\rm avg,e}} & = \mathbb{E}[ \left(\Lambda_{\rm avg,e} \left( b_{(k-1)M_{c}+l} \right) - \mu_{\Lambda_{\rm avg,e}} \right)^2 ] \\
& = \frac{1}{L} \sum_{l=1}^{L} \Big((\Lambda_{\rm avg,e} \left( b_{(k-1)M_{c}+l} \right)^2 \notag\\& -  \sum_{n=1}^{L} \Lambda_{n,e} \left( b_{(k-1)M_{c}+l}) \right).
\end{split}
\end{equation}
This suggests that the refinement benefits  come from enhancing the quality of the LLRs through their variance reduction, which shifts the LLRs with small values away from the origin.  

\subsection{Computational Complexity}
We consider the worst-case scenario 
{with} all APs to compute the computational complexity for obtaining the studied detection schemes. The major observation from the derived expressions is that decentralized detection reduces the complexity at the CPU in terms of computations since each AP locally detects its 
signal based on the available channel estimates, i.e., local detection only requires
$N \times N$ matrix inversions. On the other hand, 
$NL \times NL$ matrix inversions are required for the centralized processing scenario since all the 
combined signal is detected as a whole at the CPU 
which increases the complexity of the detectors. However, the CPU is designed to have 
high processing power 
to handle such complexity \cite{rr1,rr2}. Detailed complexity analysis for the considered detectors can be found in Table \ref{CC}. 
 It can be observed 
that the computational complexity of the decentralized and centralized detectors are of the order $\mathcal{O}(N^2LK)$ and $\mathcal{O}(N^2L^2K)$, respectively. Where $\mathcal{O}(\cdot)$ is the big O notation.
\begin{table}[h!]
\renewcommand*{\arraystretch}{1.5}
\begin{footnotesize}
\caption{Computational complexity per detector.}
\vspace{-1em}
\begin{center}
\begin{tabular}{|p{1.8cm}|p{5cm}|}
\hline
\textbf{Detector} & \textbf{{Multiplications}} \\
\hline
Decentralized-MMSE & $2N^{2}LK+2K^{2}NL+8KNL+4KL2^{M_{c}}+2M_{c}KL2^{M_{c}}+KL$  \\
\hline
Centralized-MMSE & $2N^{2}L^{2}K+8KNL+2K^{2}NL+4K2^{M_{c}}+2M_{c}K2^{M_{c}}+K$\\
\hline
Decentralized-SIC & $4N^{2}LK+2K^{2}NL+8KNL+9KL2^{M_{c}}+4M_{c}KL2^{M_{c}}+KL$  \\
\hline
Centralized-SIC & $2(NLK)^{2}+2(NL)^{2}K+K^{2}NL+5NLK+K+9K2^{M_{c}}+4M_{c}K2^{M_{c}}$ \\
\hline
Decentralized-List& $4N^{2}LK+5K^{2}NL+12KNL+9KL2^{M_{c}}+4M_{c}KL2^{M_{c}}+2KL$  \\
\hline
Centralized-List& $2(NLK)^{2}+2(NL)^{2}K+3K^{2}NL+9NLK+K+9K2^{M_{c}}+4M_{c}K2^{M_{c}}$ \\
\hline
\end{tabular}
\label{CC}
\end{center}
\end{footnotesize}
\end{table}

\subsection{Signaling Analysis}

In CF-mMIMO, the APs  
detect the signals locally or delegate the task 
fully or partially to the CPU. However, there should be a trade between the 
required front haul signaling amount and detection performance\cite{rr1}. Both the centralized and the decentralized processing require $\left(\tau_{c}-\tau_{p}\right)N$ scalars for the uplink received data and $\tau_{p}N$ complex scalars for the pilot sequences.  Additionally, the centralized processing requires  the $\frac{KLN^{2}}{2}$-dimensional spatial correlation matrix $\bm{\Omega}_{kl}$. For 
decentralized processing, the CPU does not require any statistical parameters for the spatial correlation matrix since the local channel estimates exist  at the APs. However, the CPU should 
know the $KC_{\text{leng}}L$-dimensional matrix of the LLRs in \eqref{LLR_AVG}  to compute the average, where $C_{\text{leng}}$ is the code word length. Thus, the signaling 
is summarized in Table \ref{CC2} and is analogous to the one in \cite{rr1}, with an additional knowledge of the dimension of the LLR matrix for the final decoding of the LLRs received from local processors.
\begin{table}[htb!]
\renewcommand*{\arraystretch}{1.5}
\begin{footnotesize}
\caption{Number of complex sequences to  share  via fronthaul connections, from APs to CPU.}
\vspace{-1em}
\begin{center}
\begin{tabular}{|p{2cm}|p{2.3cm}|p{2cm}|}
\hline
\textbf{Processing Scenario} & \textbf{{Each Coherence block}}& \textbf{{Statistical Parameters}} \\
\hline
Centralized &$\tau_{c}NL$ \cite{rr1} & ${KLN^{2}}/{2}$ \cite{rr1}\\
\hline
Decentralized & $\left(\tau_{c}-\tau_{p}\right)KL$ \cite{rr1} &$-$  \\
\hline
\end{tabular}
\label{CC2}
\end{center}
\end{footnotesize}
\end{table}
\subsection{Decoding Algorithm}
 The proposed detectors and the decoder iteratively exchange soft beliefs. 
The tangent function degrades the performance of the conventional sum-product algorithm (SPA), especially in the error-rate region 
\cite{r11}. Since the box-plus SPA produces less complex approximations, we use it in this paper \cite{r10,r11}. The single parity check (SPC) stage and the repetition stage are  two steps that make up the decoder.
The LLR sent from check node $(CN)_{J}$ to variable node $(VN)_{i}$ is computed as 
\begin{align}
        \Lambda_{j\longrightarrow i}=\boxplus_{i^{'}\in N(j)\diagdown i} \Lambda_{i^{'\longrightarrow j}},
\end{align}
where $\boxplus$ denotes the pairwise ``box-plus" operator given by 
\begin{align}
\Lambda_{1}\boxplus \Lambda_{2}=&\log\left ( \frac{1+e^{\Lambda_{1}+\Lambda_{2}}}{e^{\Lambda_{1}}+e^{\Lambda_{2}}} \right ),\\\notag
  =&\mathrm{sign}(\Lambda_{1})\mathrm{sign}(\Lambda_{2})\min(\left | \Lambda_{1} \right |,\left | \Lambda_{2} \right |)\notag\\&+\log\left ( 1+e^{-\left |\Lambda_{1}+\Lambda_{2}  \right |} \right )-\log\left (1+e^{-\left |\Lambda_{1}-\Lambda_{2}  \right |}  \right ).
\end{align}
The LLR from $VN_{i}$ to $CN_{j}$ is given by
\begin{align}
   \Lambda_{i\longrightarrow j}=\Lambda_{i}+\sum_{j^{'}\in N(i)\backslash j}\Lambda_{j^{'}\longrightarrow i}, 
\end{align}
where the parameter $\Lambda_{i}$ denotes the LLR at $VN_{i}$, ${j^{'}\in N(i)\backslash j}$ 
{means that }all CNs connected to $VN_{i}$ except $CN_{j}$.
 \section{Simulation Results}\label{SIM_RESULTS} 
In this section, the {bit error rate} (BER) performance of the  proposed soft detectors is presented for the CF-mMIMO settings. We use LDPC codes \cite{pegbf,vfap,memd} with codeword length $C_{\text{leng}}=256$ bits, $M=128$ parity check bits and $C_{\text{leng}}-M$ message bits. The CF-mMIMO channel exhibits high PL values due to  LS fading coefficients. Thus, the SNR 
can be expressed by
\begin{align}
    SNR=\frac{\sum_{l=1}^{L}(\mathbf{G}_{l}~\mathrm{diag}\left(\bm{\rho}\right)\mathbf{G}^{H}_{l})}{\sigma_{w}^{2} NL K}.
\end{align}
   The simulation parameters  are varied  according to table \ref{simpara},  unless stated otherwise: 
   \begin{table}[h!]
\renewcommand*{\arraystretch}{1.5}
\begin{footnotesize}
\caption{Simulation Parameters.}
\vspace{-1em}
\begin{center}
\begin{tabular}{|p{5cm}|p{2cm}|}
\hline
\textbf{Parameter} & \textbf{{Value}} \\
\hline
Codeword length ($C_{\text{leng}})$ & $256$  \\
\hline
Parity Check bits ($M$) & $128$\\
\hline
Message bits  & $C_{\text{leng}}-M$\\
\hline
Code rate $R$  & $\frac{1}{2}$\\
\hline
Threshold Euclidean distance ($d_{\text{th}}$) & $0.38$  \\
\hline
$\tau_{u}, \tau_{p}, \tau_{c}$ & $190, 10, 200$ \\
\hline
$\eta_{k}$& $100~\mathrm{mW}$  \\
\hline
Threshold for non-master AP ($\beta_{\text{th}}$)& $-20~\mathrm{dB}$ \\
 \hline
Maximum decoder iterations & $10$ \\
 \hline
 Signal power $\rho$ & $1~\mathrm{W}$  \\
 \hline
Maximum decoder iterations & $10$ \\
 \hline
 Square length of CF-mMIMO Network $D$ & $1~\mathrm{km}$  \\
   \hline
Angular standard deviation  & $15^\circ$  \\
 \hline
Noise power  & $-96~\mathrm{dBm}$  \\
 \hline
Bandwidth  & $20~MHz$  \\
  \hline
Number of channel realizations  & $10000$  \\
\hline
\end{tabular}
\label{simpara}
\end{center}
\end{footnotesize}
\end{table}

\textbf{Network setup, assumptions  and remarks}: We consider a cell-free environment with  a square of dimensions $D\times D$. The spatial correlation matrices $\bm{\Omega}_{jl}$ are assumed to be locally available at the APs and their entries are  generated using the Gaussian local scattering model \cite{rr2,RR15}
with an angular standard deviation defined in Table \ref{simpara} .
    The modulation scheme used is QPSK. The LS fading coefficients are obtained according to the  3rd Generation Partnership Project (3GPP) Urban Microcell model in \cite{rr2} given by 
  \begin{align}     \beta_{k,l}\left[\mathbf{\mathrm{dB}}\right]=-30.5-36.7\log_{10}\biggl(\frac{d_{kl}}{1 m}\biggr)+\Upsilon_{kl},
  \end{align}
 where $d_{kl}$ is the distance between the $k$-th UE and $l$-th AP,  $\Upsilon_{kl}\sim\mathcal{N}\left(0, 4^{2}\right)$ is the shadow fading.  We believe that the considered propagation and channel model are sufficiently general to allow  simple changes and assessment of line of sight (LoS), pathloss and shadowing distributions for evaluating  CF-mMIMO networks, as recommended in literature for micro cell scenarios \cite{rr2}. The simulation results are based on single antenna UEs for simplicity of analysis, the test of ideas and to allow shorter simulation time. However, this can be extended to multiple antenna UEs to 
 address the most practical systems fully. Note should also be taken that the considered scenarios and network settings in terms of codeword length, number of APs and antennas provide reasonably acceptable performances in terms of BER. However, improvements in the BER can be obtained by using longer channel codes as well increasing the number of APs and antennas in the network at the expense of increased simulation time and complexity. 
 
\begin{figure}[!htbp]
\centering
\includegraphics[width=8.5cm]{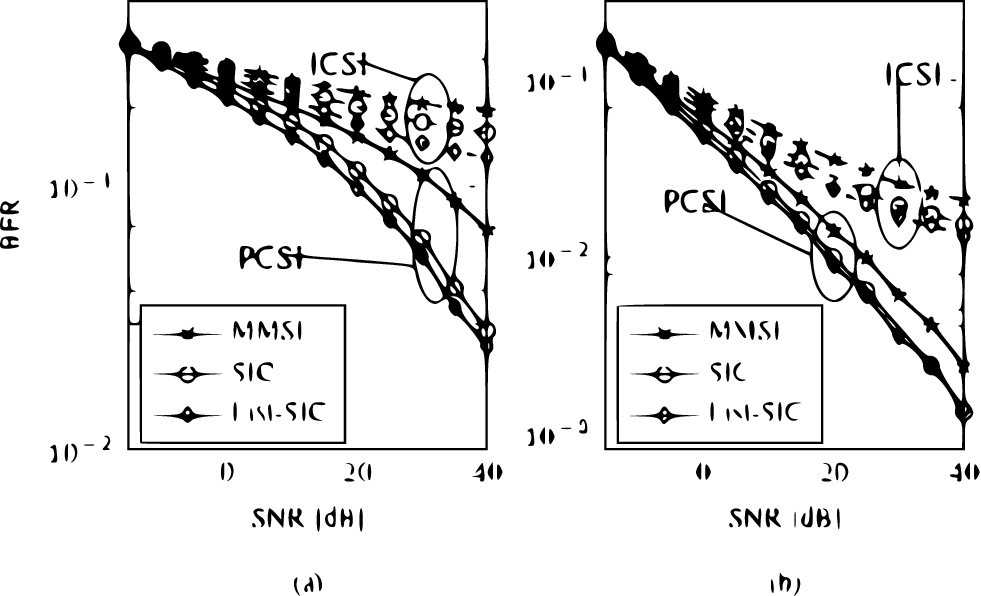}
\caption{BER versus SNR while comparing detectors for decentralized processing for $L=4$, $N=4$, $K=4$:  (a) Before LLR Refinement and (b) After LLR Refinement.}
 \label{figJ1}   
 \end{figure}

Figure \ref{figJ1} presents the BER versus the SNR for the case (a) before LLR refinement (w/o-LLR-Ref) and (b) after LLR refinement (w-LLR-Ref)  , for the studied detectors. It can be noticed that there is a significant reduction in the BER for the case with LLR refinement as compared to the scenario without LLR refinement. This performance improvement is attributed to the linear combination of the multiple streams of LLRs from the different APs which improves their reliability by shifting the LLRs with small values away from the origin. Secondly, 
some benefits arise due to the diversity of LLRs which improves the system performance, whereas for the w/o-LLR-Ref case, the hard decisions are made based on the individual APs LLRs and later an average BER is obtained. This naive approach leads to performance degradation as some APs have very unreliable estimates and hence poor LLRs. Thus, hard decisions made on LLRs from such APs compromise the entire network performance. Additionally, the figure compares the case with PCSI  and ICSI. It can be observed that the detection based on the PCSI achieves lower BERs as compared to the case with ICSI. This is because the channel estimation error and pilot contamination   degrade the network performance, 
resulting in high BERs. Another key observation is that the proposed list-based detector achieves lower BER values than the SIC detector. This is because the List-based approach can eliminate the error propagation that occurs in the conventional SIC due to the MF diversity. Note also that the linear MMSE receiver has the worst performance among the studied detectors since it does not have the $\Delta_{\text{i}}$ matrix 
used for interference cancellation. 

\begin{figure}[!htbp]
  \centering
 \includegraphics[width=8.5cm]{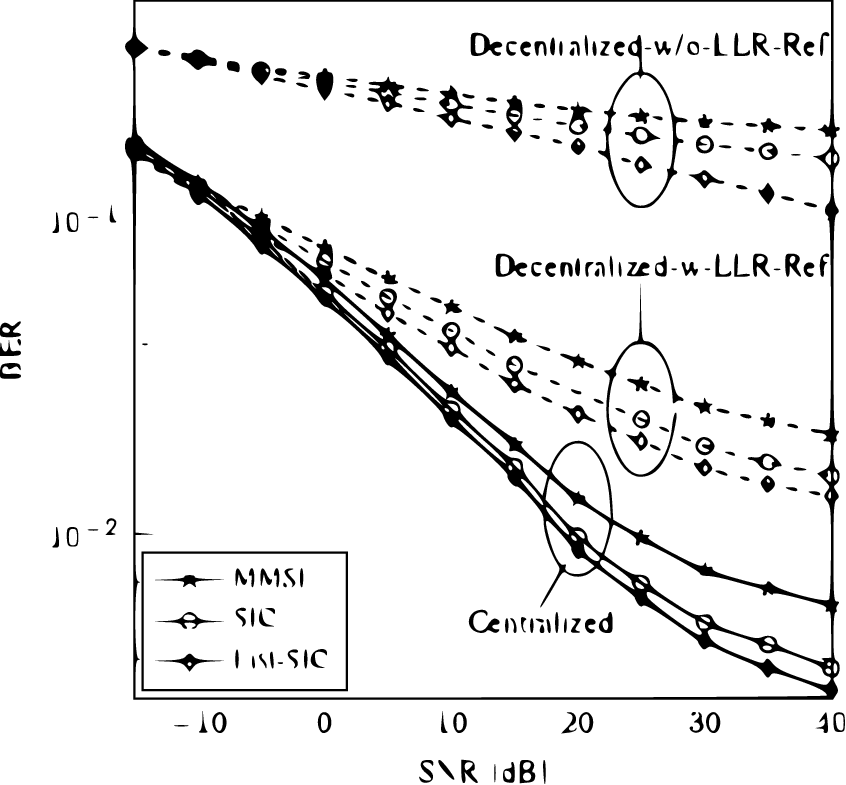}
 \caption{BER versus SNR for All APs comparing  decentralized and centralized processing for the case with imperfect CSI with $L=4$, $K=4$, $N=4$, $\mathrm{IDD}=2$. }
 \label{fig32x} 
  \end{figure}

A comparison of the centralized and decentralized processing schemes in terms of BER versus SNR is presented in Figure \ref{fig32x}. 
The results show that the case with centralized processing achieves lower BER values than the case with decentralized processing. This is because centralized processing takes  a joint detection of all the received signals into account.  Also, the case w/o-LLR-Ref achieves the worst performance since each AP performs its hard decisions locally based on the available LLRs and an average BER is obtained for the entire network, which yields a huge performance gap and degradation. The case w-LLR-Ref outperforms the standard processing scheme since it takes advantage of LLR combining. 
This yields more reliable LLRs around the mean which improves performance. This performance improvement is significant for CF-mMIMO architectures as it can yield less complex solutions in uplink detection schemes, i.e., for the decentralized processing, there is a substantial reduction in the computation complexity and the fronthaul signaling load in the network as shown in Tables \ref{CC} and \ref {CC2}, respectively. 

\begin{figure}[t!]
  \centering
\includegraphics[width=8.5cm]{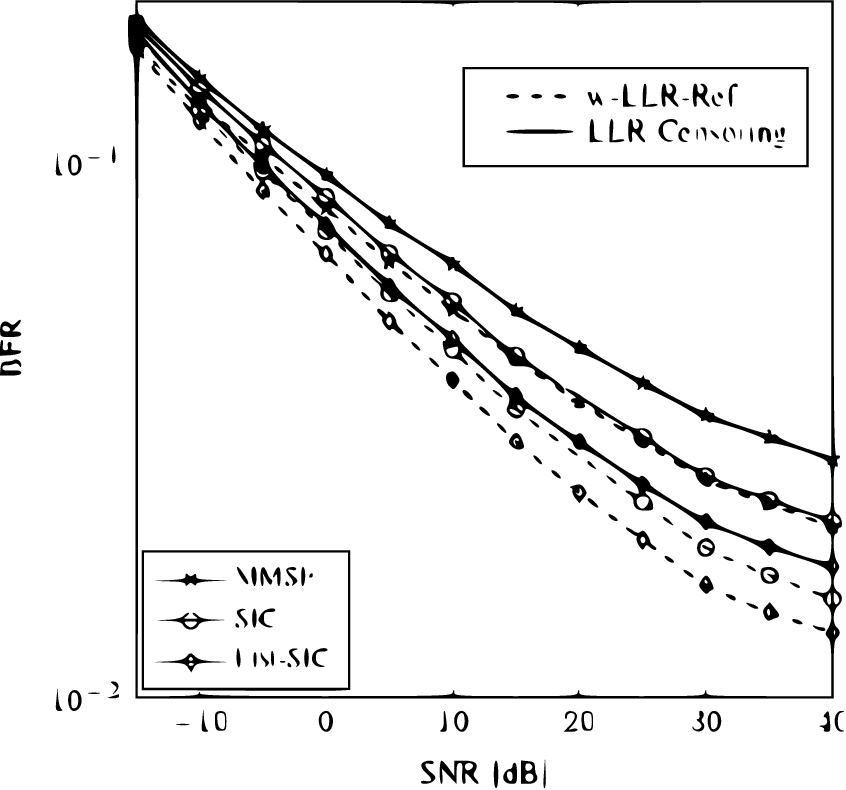}
 \caption{BER versus SNR for All APs comparing  LLR Censoring and LLR Refinement for decentralized processing  for the case with imperfect CSI with $L=4$, $K=4$, $N=4$, $\mathrm{IDD}=2$. }
 \label{fig32} 
\end{figure}
  
 Figure \ref{fig32} shows the BER versus SNR for the decentralized processing cases using w-LLR-Ref and LLR-Censoring. It is clear  that the case using LLR refinement has lower BER than the one using LLR censoring. 
 The UE can only be decoded at the AP, 
 achieving the highest mean absolute value when LLR censoring is used. In contrast, LLR refinement enhances performance by performing a linear combination of the LLRs from all APs.  Nonetheless, censoring LLRs prevents the redundant processing of the LLRs. The only difficulty that might arise is a slight increase in hardware complexity of the receiver design since the CPU must constantly scan all the APs to identify the one that offers the largest absolute value of LLRs to a specific UE. However, the CPU is usually designed with a strong computing power and thus it can handle such complexity. One could confuse the proposed LLR censoring and refinement to be analogous to the selection combining and maximal ratio combining used in diversity analysis. However, the former leverages the distributed computation of LLRs from each AP and therefore it should not be confused with the latter schemes.

\begin{figure}[!htbp]
 \centering
 \includegraphics[width=8.5cm]{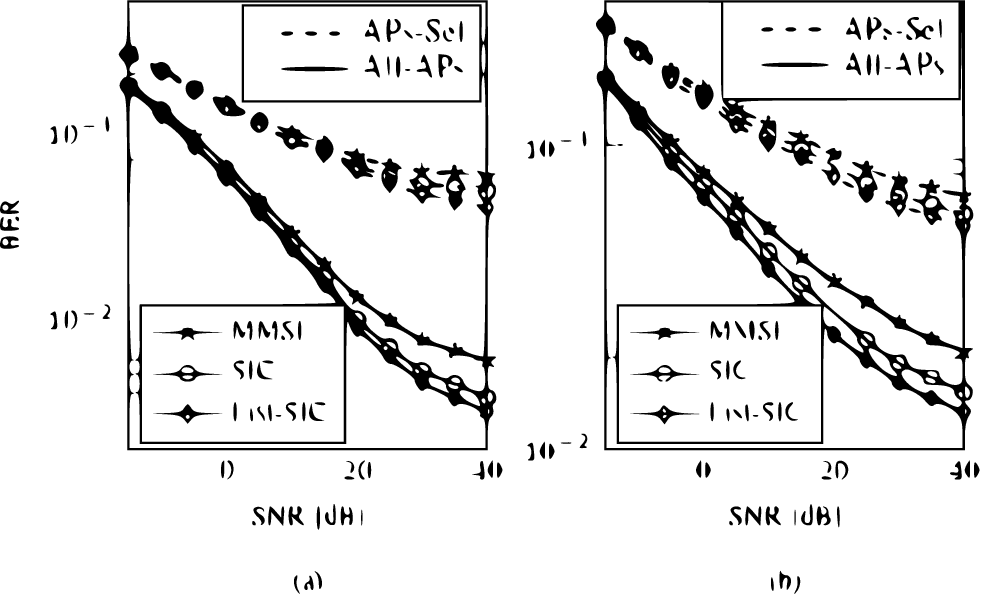}
 \caption{BER versus SNR for a case that uses All APs and a case that uses APs-Sel for $L=4$, $N=4$, $K=4$:  (a) Centralized Processing and (b) Decentralized Processing}.
  \label{figJ3}   
  \end{figure}

Figure \ref{figJ3} plots the BER versus SNR for the case that the detectors use all APs (All-APs) and the case that uses APs selection (APs-Sel) with (a) centralized and (b) decentralized processing schemes. It can be observed that for both processing levels, the system that uses All-APs achieves lower BER values as compared to the one that selects the APs. This is 
because selecting APs reduces the number of antennas in the network, 
distorting the performance. However, APs selection reduces the signaling load, 
 making the network more scalable and practical. Moreover, the distributed location, the delay spread of the APs and the associated signal propagation latency will limit the APs involved in cell-free MIMO systems. Therefore, APs selection techniques are key to reducing fronthaul
signaling, computational costs and latency. 

\begin{figure}[!htbp]
\centering
\includegraphics[width=8.5cm]{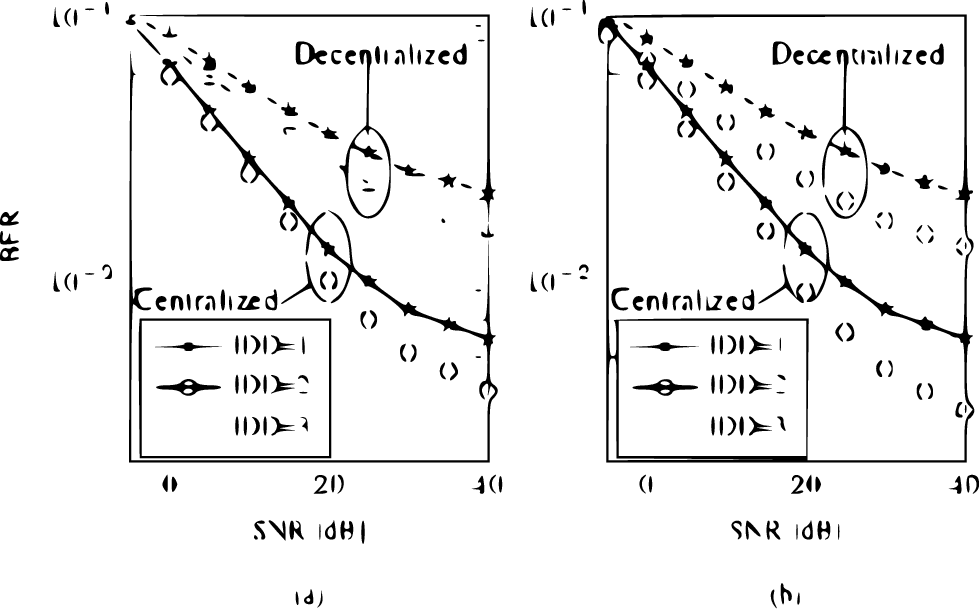}
\caption{BER versus SNR while varying number of IDD iterations  for $L=4$, $N=4$, $K=4$:  (a) SIC and (b) List-SIC.}
 \label{figJ4}   
 \end{figure}

The BER versus SNR of the detectors while varying the number of outer iterations for the detectors is presented in Figure \ref{figJ4}.  From the curves, it can be noticed that increasing the number of iterations reduces the BER. Specifically, for both centralized and decentralized (a) SIC and (b) List-SIC, there is a significant performance improvement when the number of iterations is increased from $1$ to $2$ iterations for both detectors.  However, after the third iteration, the performance benefits are marginal. 

\begin{figure}[!htbp]
\centering
\includegraphics[width=8.5cm]{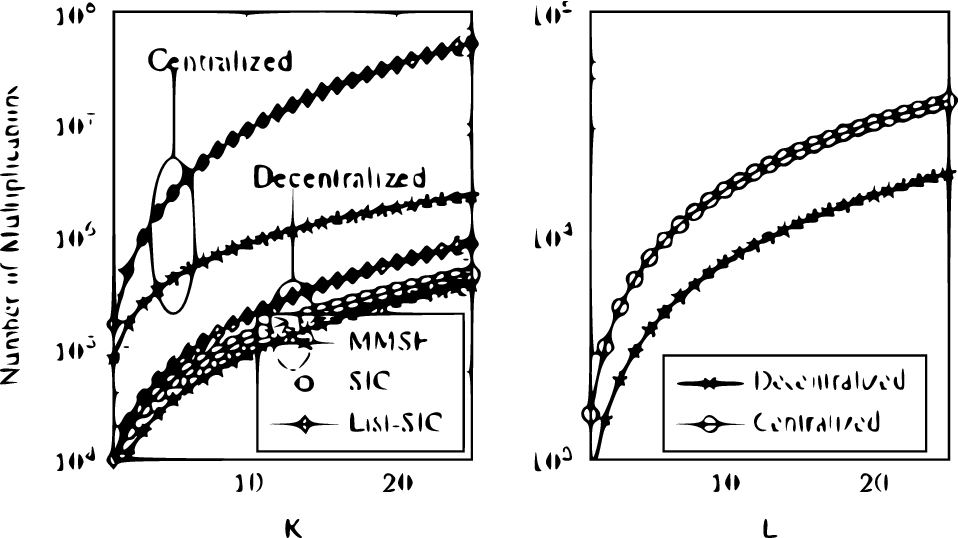}
\caption{Number of multiplications versus number of UE $K$  and number of APs $L$ (a) Computational complexity for $L=50$, $M_{c}=2$, $N=4$ and (b) Signaling load for $K=4$,  $N=8$.}
\label{fig36}   
\end{figure}

Figure \ref{fig36} (a) plots the computational complexity versus the number of UEs $K$ while considering the studied  detectors for centralized and decentralized processing. It can be observed that the linear MMSE detectors have the lowest computation complexity for both cases. The list-based detectors have slightly higher computation complexity than the SIC-based detectors. It is worth mentioning that the differences in computational complexity between SIC and List-SIC are marginal, whereas centralized detectors have higher computation cost than decentralized detectors.   

The signaling load is given by the number of complex scalars 
that must be exchanged in the network, which is shown  in Figure \ref{fig36} (b) for the centralized and decentralized CF-mMIMO setups. From the curves it can be noticed that decentralized processing requires less signaling between the APs and CPU 
than the centralized setup. The decentralized processing requires knowledge of the $KC_{\text{leng}}L$-dimensional matrix of the LLRs from all APs to perform the final processing. Nevertheless, decentralized schemes greatly reduce the required signaling in the network and can achieve close performance to that of centralized processing while using LLR refinement.
\FloatBarrier.
\section{Conclusions}\label{Concl}
In this paper, an IDD scheme using LDPC codes has been devised with APs selection for centralized and decentralized CF-mMIMO architectures. In particular, we have proposed low-complexity interference mitigation techniques including a list-based detector that uses MMSE receive filter to improve the performance. New closed-form expressions for the MMSE-soft-IC detectors have been derived for both the centralized and decentralized implementations, taking  the channel estimation errors and APs selection into account. The performance of the proposed list-based detector is compared with other baseline detectors, such as the soft linear MMSE, MMSE-SIC, and the results show that the list-based detector yields low BER values compared to the other detectors. We also proposed LLR refinement strategies 
 based on combining and censoring LLRs. The results have shown that LLR refinement strategies obtained lower BER values 
 than standard processing. The proposed scalable APs-Sel APs selection scheme based on LLSF coefficients can reduce the signaling load between APs and the CPU, resulting in a trade-off between performance, signaling load and network feasibility. 
\appendix
\subsection{Derivation of the Proposed Centralized Detector}\label{Deriv_cent}
We start our analysis by expressing the conditional expectation on the RHS of \eqref{funp2lp} as
\begin{align}\label{obj1}
    F= \mathbb{E}\biggl\{||\tilde{s}_{k}-s_{k}||^{2}\mid \hat{\mathbf{G}}\biggr\}=\mathbb{E}\biggl\{\left ( \tilde{s}_{k}-s_{k} \right )\left ( \tilde{s}_{k}-s_{k}\right )^{*}\mid \hat{\mathbf{G}}\biggr\}.
\end{align}
Substituting \eqref{funAPS1CENT} into \eqref{obj1} yields
\begin{align}\label{obj2v}
&F =\mathbf{w}^{H}_{k}{\mathbf{D}}_{k}\mathbb{E}\left \{ \left ( \mathbf{y}-\hat{\mathbf{G}}_{\text{i}}\bar{\mathbf{s}_{\text{i}}} \right )\left ( \mathbf{y}^{H}-\bar{\mathbf{s}}^{H}_{\text{i}}\hat{\mathbf{G}}^{H}_{\text{i}} \right ) \right \}{\mathbf{D}}^{H}_{k}\mathbf{w}_{k}\notag\\&-\mathbf{w}^{H}_{k}{\mathbf{D}}_{k}\mathbb{E}\left \{ \left ( \mathbf{y}-\hat{\mathbf{G}}_{\text{i}}\bar{\mathbf{s}} _{\text{i}}\right )s^{*}_{k}\right \}-\mathbb{E}\left \{s_{k} \left ( \mathbf{y}^{H}-\bar{\mathbf{s}^{H}_{\text{i}}}\hat{\mathbf{G}}^{H}_{\text{i}} \right )\right \}\notag\\&\times{\mathbf{D}}^{H}_{k}\mathbf{w}_{k}+\mathbb{E}\left \{ s_{k}s^{*}_{k} \right \}.
\end{align}
Further simplification of \eqref{obj2v} can be done by letting $\mathbf{y}_{R}=\mathbf{y}-\hat{\mathbf{G}}_{\text{i}}\bar{\mathbf{s}_{\text{i}}}$. Thus, \eqref{obj2v} can be re-written as
\begin{align}\label{obj3}
 F &=\mathbf{w}^{H}_{k}{\mathbf{D}}_{k}\mathbb{E}\left \{ \mathbf{y}_{R}\mathbf{y}^{H}_{R} \right \}{\mathbf{D}}^{H}_{k}\mathbf{w}_{k}-\mathbf{w}^{H}_{k}{\mathbf{D}}_{k}\mathbb{E}\left \{ \mathbf{y}_{R}s^{*}_{k}\right \}\notag\\&-\mathbb{E}\left \{s_{k} \mathbf{y}^{H}_{R}\right \}{\mathbf{D}}^{H}_{k}\mathbf{w}_{k}+\mathbb{E}\left \{ s_{k}s^{*}_{k} \right \}.
\end{align}
Differentiating \eqref{obj3} with respect to w.r.t $\mathbf{w}^{H}_{k}$  we obtain
\begin{align}\label{obj5}
    \frac{\partial F}{\partial \mathbf{\mathbf{w}^{H}_{k}}}&={\mathbf{D}}_{k}\mathbb{E}\left \{ \mathbf{y}_{R}\mathbf{y}^{H}_{R} \right \}{\mathbf{D}}^{H}_{k}\mathbf{w}_{k}-{\mathbf{D}}_{k}\mathbb{E}\left \{ \mathbf{y}_{R}s^{*}_{k}\right \}.
\end{align}
The optimal MMSE filter is obtained by  equating \eqref{obj5} to ${\bf 0}$.  Thus, the optimal MMSE filter $\mathbf{w}_{k}$ is given by
\begin{align}\label{obj6}
   {\mathbf{D}}_{k}\mathbb{E}\left \{ \mathbf{y}_{R}\mathbf{y}^{H}_{R}\right \}{\mathbf{D}}^{H}_{k}\mathbf{w}_{k}-{\mathbf{D}}_{k}\mathbb{E}\left \{ \mathbf{y}_{R}s^{*}_{k} \right \}=0.
\end{align}
The reader can confirm that \eqref{obj6} is the same as \eqref{objxx}. By making $\mathbf{w}_{k}$ the subject of \eqref{obj6}, we obtain \eqref{fcentC}. The terms $ \mathbb{E}\{\mathbf{y}_{R}{s}^{*}_{k}\}$ and $  \mathbb{E}\{\mathbf{y}_{R}\mathbf{y}^{H}_{R}\}$ are given by \eqref{w1C} and \eqref{w2C}, where $\mathbb{E}\left \{ s_{m}s^{*}_{m} \right \}=\left | s_{m} \right |^{2}+\sigma^{2}_{m}$, $\mathbb{E}\left \{ \tilde{\mathbf{g}}_{m}\tilde{\mathbf{g}}^{H}_{m} \right \}=\mathbf{C}_{m}$, $\mathbb{E}\left \{ \mathbf{n}\mathbf{n}^{H} \right \}=\sigma^{2}\mathbf{I}_{NL}$, $\mathbb{E}\left \{ s_{k}s^{H}_{k} \right \}=\rho_{k}$, obtained after assuming statistical independence between each term in the RHS of \eqref{receiv_CPU_signalAPSEL} and using the orthogonality principle \cite{RR14}.
By substituting \eqref{w1C} and \eqref{w2C} into \eqref{fcentC}, we arrive at the centralized MMSE filter given by \eqref{det_2}.
\subsection{Derivation of the Proposed Decentralized Detector}
The derivation of the proposed local MMSE filter is  similar to that of Appendix \ref{Deriv_cent}. 
The expectation on the R.H.S of \eqref{funp2lp11} can be expressed as
\begin{align}\label{obj1x}
    F_{2}&= \mathbb{E}\biggl\{||\tilde{s}_{kl}-s_{k}||^{2}\mid \hat{\mathbf{G}}_{l}\biggr\}\notag\\&=\mathbb{E}\biggl\{\left ( \tilde{s}_{kl}-s_{k} \right )\left ( \tilde{s}_{kl}-s_{k}\right )^{*}\mid \hat{\mathbf{G}}_{l}\biggr\}.
\end{align}
By substituting \eqref{funAPS1} into \eqref{obj1x}  we obtain
\begin{align}\label{decx1}
    F_{2}&=\mathbb{E}\left \{\left (   \mathbf{w}^{H}_{kl}{\mathbf{D}}_{kl}\mathbf{y}_{Rl}-s_{k}\right )\left (   \mathbf{w}^{H}_{kl}{\mathbf{D}}_{kl}\mathbf{y}_{Rl}-s_{k}\right )^{*}\right \}.
\end{align}
The term $\mathbf{y}_{Rl}$ is the residue signal obtained after soft-IC and substituting for $y_{l}$ in the term in brackets of \eqref{funAPS1}, we get
\begin{align}   \mathbf{y}_{Rl}=\hat{\mathbf{g}}_{kl}s_{k}+\hat{\mathbf{G}}_{\text{i}l}\left ( \mathbf{s}_{\text{i}}-\bar{\mathbf{s}}_{\text{i}} \right )+\sum_{m=1}^{K}\tilde{\mathbf{g}}_{ml}s_{m}+\mathbf{n}_{l}.
\end{align}
After some mathematical and algebraic manipulations, \eqref{decx1} can be re-written as
\begin{align}\label{f2n2}
F_{2}&=\mathbf{w}^{H}_{kl}{\mathbf{D}}_{kl}\mathbb{E}\left \{ \mathbf{y}_{Rl}\mathbf{y}^{H}_{Rl} \right \}{\mathbf{D}}^{H}_{kl}\mathbf{w}_{kl}-\mathbf{w}^{H}_{kl}{\mathbf{D}}_{kl}\mathbb{E}\left \{ \mathbf{y}_{Rl}s^{*}_{k} \right \}\notag\\&-\mathbb{E}\left\{s_{k}\mathbf{y}^{H}_{Rl}\right\}{\mathbf{D}}^{H}_{kl}\mathbf{w}_{kl}+\mathbb{E}\left \{ s_{k}s^{*}_{k} \right \}
\end{align}
We take the first derivative of \eqref{f2n2} w.r.t $\mathbf{w}^{H}_{kl}$ to arrive at
\begin{align}\label{f2n3}
     \frac{\partial F_{2}}{\partial {\mathbf{w}}^{H}_{kl}}&={\mathbf{D}}_{kl}\mathbb{E}\left \{ \mathbf{y}_{Rl}\mathbf{y}^{H}_{Rl} \right \}{\mathbf{D}}^{H}_{kl}\mathbf{w}_{kl}-{\mathbf{D}}_{kl}\mathbb{E}\left \{ \mathbf{y}_{Rl}s^{*}_{k} \right \}.
\end{align}
 After equating the resulting expression in \eqref{f2n3} to $0$, we obtain
\begin{align}\label{f2n4}
    {\mathbf{D}}_{kl}\mathbb{E}\left \{ \mathbf{y}_{Rl}\mathbf{y}^{H}_{Rl} \right \}{\mathbf{D}}^{H}_{kl}\mathbf{w}_{kl}-{\mathbf{D}}_{kl}\mathbb{E}\left \{ \mathbf{y}_{Rl}s^{*}_{k} \right \}=0.
\end{align}
The optimal local MMSE filter $\mathbf{w}_{kl}$ can be obtained from \eqref{f2n4}.
The terms $\mathbb{E}\left\{\mathbf{y}_{Rl}\mathbf{y}^{H}_{Rl}\right\}$ and $\mathbb{E}\left\{\mathbf{y}_{Rl}s^{*}_{k}\right\}$ can be obtained from \eqref{w1d} and \eqref{w2d}, respectively, where the terms $\mathbb{E}\left\{\tilde{\mathbf{g}}_{ml}\tilde{\mathbf{g}}^{H}_{ml}\right\}=\mathbf{ C}_{ml}$ and $\mathbb{E}\left\{\mathbf{n}_{l}\mathbf{n}^{H}_{l}\right\}=\sigma^{2}\mathbf{I}_{N}$, by taking assumptions similar to those in Subsection A.
\subsection{Derivation of the Soft Demapper Parameters for Centralized and Decentralized Processing}
We start the proof by making some assumptions on the output of the MMSE filter to be a Gaussian approximation. The optimal soft bit metric which takes into account the channel estimation error and APs-Sel can be derived as below.
Let $k$ denote the desired UE which minimizes the mean square error (MSE). Then, \eqref{funAPS1CENT} can be expressed as
\begin{align}\label{soft_demap1}   \tilde{s}_{k}=\mathbf{w}^{H}_{k}{\mathbf{D}}_{k}\mathbf{y}-\mathbf{w}^{H}_{k}{\mathbf{D}}_{k}\hat{\mathbf{G}}_{i}\bar{\mathbf{s}}_{\text{i}}.
\end{align}
By substituting \eqref{receiv_CPU_signalAPSEL} into \eqref{soft_demap1} we obtain
\begin{align}\label{soft_demap2}  \tilde{s}_{k}&=\mathbf{w}^{H}_{k}{\mathbf{D}}_{k}\hat{\mathbf{g}}_{k}s_{k}+\mathbf{w}^{H}_{k}{\mathbf{D}}_{k}\hat{\mathbf{G}}_{\text{i}}\left ( \mathbf{s}_{\text{i}}-\bar{\mathbf{s}}_{\text{i}} \right )+\mathbf{w}^{H}_{k}{\mathbf{D}}_{k}\sum_{m=1}^{K}\tilde{\mathbf{g}}_{m}s_{m}\notag\\&+\mathbf{w}^{H}_{k}{\mathbf{D}}_{k}\mathbf{n}.
\end{align}
By comparing \eqref{soft_demap2} with \eqref{idd1}, it can be observed that 
\begin{align}\label{soft_demap3}   \omega_{k}=\mathbf{w}^{H}_{k}\tilde{\mathbf{D}}_{k}\hat{\mathbf{g}}_{k},
\end{align} and the interference-plus-noise term is given by 
\begin{align}\label{soft_demap4}   z_{k}=\mathbf{w}^{H}_{k}{\mathbf{D}}_{k}\hat{\mathbf{G}}_{\text{i}}\left ( \mathbf{s}_{\text{i}}-\bar{\mathbf{s}}_{\text{i}} \right )+\mathbf{w}^{H}_{k}{\mathbf{D}}_{k}\sum_{m=1}^{K}\tilde{\mathbf{g}}_{m}s_{m}+\mathbf{w}^{H}_{k}{\mathbf{D}}_{k}\mathbf{n}.
\end{align}
By assuming that $z_{k}$ is a Gaussian random variable \cite{r3} and assuming statistical independence of each term of \eqref{soft_demap4}, the variance $\kappa^{2}=\mathbb{E}\left\{\mid u_{k}-\omega_{k}s_{k}\mid^{2}\right\}=\mathbb{E}\left\{z_{k}z^{*}_{k}\right\}$ of $z_{k}$ is given by
\begin{align}\label{sdemapcent2}
\kappa^{2}=\mathbf{w}^{H}_{k}{\mathbf{D}}_{k}\left(\hat{\mathbf{G}}_{\text{i}}\Delta_{\text{i}}\hat{\mathbf{G}}^{H}_{\text{i}}+\sum_{m=1}^{K}\rho_{m}\mathbf{C}_{m}+\sigma^{2}\mathbf{I}_{NL} \right ){\mathbf{D}}^{H}_{k}\mathbf{w}_{k}.
\end{align}
By substituting \eqref{soft_demap3} and \eqref{sdemapcent2} into \eqref{llfn}, the soft beliefs can be obtained in each subsequent iteration.
Using similar procedures, the parameter $\kappa^{2}_{l}$ for the decentralized processing can be obtained as given in \eqref{sdemapdeccent}.


\vspace{-0.1em}
\begin{thebibliography}{00}

\bibitem{rr1}
E.~Bj{\"o}rnson and L. Sanguinetti,'' Making cell-free massive MIMO
competitive with MMSE processing and centralized implementation'',
\emph{ IEEE Trans. Wireless Commun.}, vol. 19, no. 1, pp. 77–90, Jan. 2020. 

 \bibitem{rr2}
E.~Bj{\"o}rnson and L. Sanguinetti, '' Scalable Cell-Free Massive MIMO Systems'',
\emph{ IEEE Trans. Wireless Commun.},  vol. 68, no. 7, pp. 4247-4261, Jul. 2020.

\bibitem{r9}
 H. Q. Ngo, A. Ashikhmin, H. Yang, E. G. Larsson and T. L. Marzetta, "Cell-Free Massive MIMO Versus Small Cells," \emph{IEEE Trans. Wireless Commun.}, vol. 16, no. 3, pp. 1834-1850, Mar. 2017.
 
 \bibitem{rr3}
 Z.~ Chen and E.~Bj{\"o}rnson,
 ''Channel hardening and favorable propagation in cell-free massive MIMO with stochastic geometry'', \emph{ IEEE Trans. Wireless Commun.}, vol. 66, no. 11, pp. 5205-5219, Nov. 2018.

 \bibitem{rr4}
H. T. Dao and S. Kim, '' Effective channel gain-based access point
selection in cell-free massive MIMO systems'',\emph{ IEEE Access}, vol. 8, pp. 108127–108132, June 2020. 

 \bibitem{rr5}
 H. Q. Ngo, L.-N. Tran, T. Q. Duong, M. Matthaiou, and E. G. Larsson, '' On the total energy efficiency of cell-free massive MIMO'', \emph{ IEEE
Trans. Green Commun. Netw}, vol. 2, no. 1, pp. 25–39, Mar. 2018.

 \bibitem{rr61xx}
S. Mashdour, R. C. de Lamare and J. P. S. H. Lima, '' Enhanced Subset Greedy Multiuser Scheduling in Clustered Cell-Free Massive MIMO Systems'',
\emph{ IEEE Commun. Lett}, vol. 27, no. 2, pp. 610-614, Feb. 2023. 
 
\bibitem{rr62xx}
S. Mashdour, R. C. de Lamare, A. Schmeink and J. P. S. H. Lima,  '' MMSE-Based Resource Allocation for Clustered Cell-Free Massive MIMO Networks'',
\emph{  WSA \& SCC 2023; 26th International ITG Workshop on Smart Antennas and 13th Conference on Systems, Communications, and Coding}, Braunschweig, Germany, 2023, pp. 1-6.

 \bibitem{rr6}
Shakya, I.L., Ali, F.H.,  '' Joint access point selection and interference cancellation for cell-free massive MIMO'',
\emph{ IEEE Commun. Lett}, vol. 25, no. 4, pp. 1313--1317, Apr. 2021. 
 \bibitem{rrv6}
Y. Li, Q. Lin, Y.-F. Liu, B. Ai, and Y.-C. Wu,
'' Asynchronous activity detection for cell-free massive MIMO: From centralized to distributed algorithm'', \emph{ IEEE Trans. Commun.}, pp. 1--1, Oct. 2022.

 \bibitem{r1}
 X. Wang and H.~V.~Poor.
	    '' Iterative (turbo) soft interference cancellation and decoding for coded CDMA'',
\emph{ IEEE Trans. Commun.}, vol. 47, no. 7, Jul. 1999. 
\bibitem{r2}
 B. Xiao, K. Xiao, S. Zhang, Z. Chen, B. Xia and H. Liu., ''Iterative detection and decoding for SCMA systems with LDPC codes'' \emph{in~Proc. Int. Conf. on Wireless Commun. and Signal Process. (WCSP)}, Nanjing, China, 15-17 Oct. 2015,  pp. 1-7, 19-22.
\bibitem{r3}
A.~Matache, C.~ Jones and R.~D.~Wesel, ''Reduced complexity MIMO detectors for LDPC coded systems'', \emph{in~Proc.  IEEE Military Commun. Conf.}, Monterey, CA, USA, 31 Oct.-3 Nov. 2004, pp. 1073-1079.
\bibitem{r3v}
F.~Sagheer, F.~ Lehmann, and A.~O.~Berthet. 'Low-Complexity Dynamic Channel Estimation in
Multi-Antenna Grant-Free NOMA'', \emph{in~Proc.  2022 IEEE 95th Veh. Techno. Conf.: (VTC2022-Spring).}, Helsinki, Finland, 19-22 June 2022, pp. 1-7.

\bibitem{spa}
R. C. De Lamare and R. Sampaio-Neto, "Minimum Mean-Squared Error Iterative Successive Parallel Arbitrated Decision Feedback Detectors for DS-CDMA Systems," in \emph{IEEE Trans. Commun.}, vol. 56, no. 5, pp. 778-789, May 2008.

\bibitem{spa2}
 R.~B. Di~Renna, and R. C. De Lamare, "Joint Channel Estimation, Activity Detection and Data Decoding Based on Dynamic Message-Scheduling Strategies for mMTC," in \emph{IEEE Trans. Commun.}, vol. 70, no. 4, pp. 2464 - 2479, April 2022.
 
 \bibitem{spa1}
H. Li, Y. Dong, C. Gong, X. Wang and X. Dai, "Gaussian Message Passing Detection With Constant Front-Haul Signaling for Cell-Free Massive MIMO," 
\emph{ IEEE Trans. Veh. Technol.}, pp. 1--6, Nov. 2022.

\bibitem{spa3}
 R.~B. Di~Renna, and R. C. De Lamare, "Iterative List Detection and Decoding for Massive Machine-Type Communications," in \emph{IEEE Trans. Commun.}, vol. 68, no. 10, pp. 6276 - 6288, Oct. 2020.
 
\bibitem{spa4}
Z.~Shao, R.~C.~de Lamare and L.~T.~N.~Landau, "Dynamic oversampling for 1-bit ADCs in large-scale multiple-antenna systems," in \emph{IEEE Trans. Commun.}, vol. 69, no. 5, pp. 3423 - 3435, May 2021. 

\bibitem{spa5}
H. T. Nguyen, D. A. Hoang, H. T. Bui, H. N. Dang and T. V. Nguyen, "Large-Scale MIMO Communications With Low-Resolution ADCs Using 16-ary QAM and Protograph LDPC Codes,"," in "\emph{Proc. 9th IEEE  Int. Conf. on Commun. and Electron. (ICCE),} Nha Trang, Vietnam, 27-29 July 2022, pp. 43-47.

 \bibitem{r4}
 S. Jing et al., ''Joint Detection and Decoding of Polar-Coded SCMA Systems''  \emph{ in Proc.~9th Int. Conf. on Wireless Commun. and Signal Process. . (WCSP)}, Nanjing, China, 11-13 Oct. 2017, pp. 1-6.
 
\bibitem{r5}  P.~Li, R.~C.~de Lamare and R.~Fa,
 ''Multiple Feedback Successive Interference Cancellation Detection for Multiuser MIMO Systems'', \emph{IEEE Trans. Commun.}, vol. 10, no. 8, pp. 2434 - 2439, Jun. 2011.

\bibitem{mbdf+o}
 R.~C.~de Lamare, ''Adaptive and Iterative Multi-Branch MMSE Decision Feedback Detection Algorithms for Multi-Antenna Systems'', 
 \emph{IEEE Trans. Commun.}, vol. 12, no. 10, pp.~5294 - 5308, Sept.~2013.

\bibitem{r7}
 C. D’Andrea and E. G. Larsson, "Improving Cell-Free Massive MIMO by Local Per-Bit Soft Detection," \emph{ IEEE Commun. Lett.}, vol. 25, no. 7, pp. 2400-2404, Apr. 2021.

\bibitem{r77}
H. He, H. Wang, X. Yu, J. Zhang, S. H. Song and K. B. Letaief, "Distributed Expectation Propagation Detection for Cell-Free Massive MIMO," \emph{ in Proc. IEEE Glob. Commun. Conf., }  Madrid, Spain, 2021, pp. 01-06.

\bibitem{r8}
A. G. D. Uchoa, C. T. Healy and R. C. de Lamare, "Iterative Detection and Decoding Algorithms for MIMO Systems in Block-Fading Channels Using LDPC Codes," in IEEE Transactions on Vehicular Technology, vol. 65, no. 4, pp. 2735-2741, April 2016.

\bibitem{RR10}
A. Krebs, M. Joham and W. Utschick, "Comparative performance evaluation of error regularized Turbo-MIMO MMSE-SIC detectors in Gaussian channels.,"\emph{ in Proc.   EEE Int. Conf. on Acoust. Speech and Signal Process. (ICASSP)}, South Brisbane, QLD, Australia, 19-24 Apr. 2015, pp. 2984-2988.

\bibitem{RR11x}
Z. H. Shaik, E. Björnson, and E. G. Larsson, "Distributed computation of
a posteriori bit likelihood ratios in cell-free massive MIMO,"\emph{  in Proc. Eur.
Signal Process. Conf.},  Dublin, Ireland, 23-27 Aug. 2021, pp. 935-939.

\bibitem{RR11}
Heunchul Lee and Inkyu Lee, "New approach for coded layered space-time OFDM systems,"\emph{ in Proc.   IEEE Int. Conf. on Commun. (ICC)},  Seoul, Korea (South), 16-20 May 2005, pp. 608-612.

\bibitem{RR12}
Henuchul Lee, Byeongsi Lee and Inkyu Lee, "Iterative detection and decoding with an improved V-BLAST for MIMO-OFDM systems.,"\emph{ in IEEE J. Sel. Areas Commun.}, vol. 24, no. 3, pp. 504-513, Mar. 2006.

\bibitem{r14}
Z.~Shao, R.~C.~de Lamare and L.~T.~N.~Landau, "Iterative Detection and Decoding for Large-Scale Multiple-Antenna Systems With 1-Bit ADCs,"\emph{ IEEE Wireless Commun. Lett.}, vol.~7, no.~3, pp.~476-479, Jun. 2018.

\bibitem{r10}
T.~Ssettumba, R.~B. Di~Renna, L.~T.~N. Landau and R.~C.~de~Lamare, "Iterative Detection and Decoding for Cell-Free Massive MIMO Using LDPC Codes,"\emph{ in Proc.   XL Brazilian Symp. on Telecommun. and Signal Process. (SBrT) }, Sta. Rita. do Sapucai, MG.,Brasil, 25-28 Sept.~2022,  pp. 1-5.

\bibitem{r11}
T.~Ssettumba, R.~B. Di~Renna, L.~T.~N. Landau and R.~C.~de~Lamare, "List-Based Detector and Access Points Selection in Cell-Free Massive MIMO Using LDPC Codes.,"\emph{ in Proc.   Int. Symp. on Wireless Commun. Syst. (ISWCS)}, Hangzhou, China, 19-22 Oct.~2022, p. 1-6. 

\bibitem{RR13}
R.~B. Di~Renna, and R.~C.~de~Lamare, "Adaptive LLR-based APs selection for Grant-Free Random Access in Cell-Free Massive MIMO."\emph{ in Proc.     IEEE GC Wkshps}, Rio de Janeiro, Brazil, 04--08 Dec.~2022, pp. 196-201, doi: 10.1109/GCWkshps56602.2022.10008736.

\bibitem{RR14}
J. G. Proakis, "\emph{ Digital Communications.}, Fourth Edition, McGraw-Hill Series in Electrical and Computer Engineering, 2001.

\bibitem{RR15}
E.~ Björnson, J.~ Hoydis, and L.~ Sanguinetti, "Massive MIMO networks:
Spectral, energy, and hardware efficiency,"\emph{ in Foundations and Trends
in Signal Process.}, vol.~11, no.~3-4, Nov. 2017, pp.~154-655.

\bibitem{itermmsecf}
Palhares, V.M.T., de Lamare, R.C., Flores, A.R. and Landau, L.T.N. (2020), Iterative AP selection, MMSE precoding and power allocation in cell-free massive MIMO systems. IET Communications, 14: 3996-4006.

\bibitem{rmmsecf}
V. M. T. Palhares, A. R. Flores and R. C. de Lamare, "Robust MMSE Precoding and Power Allocation for Cell-Free Massive MIMO Systems," in IEEE Transactions on Vehicular Technology, vol. 70, no. 5, pp. 5115-5120, May 2021.

\bibitem{cesg}
S. Mashdour, R. C. de Lamare and J. P. S. H. Lima, "Enhanced Subset Greedy Multiuser Scheduling in Clustered Cell-Free Massive MIMO Systems," in IEEE Communications Letters, vol. 27, no. 2, pp. 610-614, Feb. 2023.

\bibitem{rscf}
A. R. Flores, R. C. de Lamare and K. V. Mishra, "Clustered Cell-Free Multi-User Multiple-Antenna Systems With Rate-Splitting: Precoder Design and Power Allocation," in IEEE Transactions on Communications, vol. 71, no. 10, pp. 5920-5934, Oct. 2023.

\bibitem{clust&sched}
S. Mashdour, S. Salehi, R. C. de Lamare, A. Schmeink and J. P. S. H. Lima, "Clustering and Scheduling With Fairness Based On Information Rates for Cell-Free MIMO Networks," IEEE Wireless Communications Letters, 2024.


\bibitem{mmimo}
R. C. de Lamare, "Massive MIMO systems: Signal processing challenges and future trends," in URSI Radio Science Bulletin, vol. 2013, no. 347, pp. 8-20, Dec. 2013

\bibitem{wence}
W. Zhang et al., "Large-Scale Antenna Systems With UL/DL Hardware Mismatch: Achievable Rates Analysis and Calibration," in IEEE Transactions on Communications, vol. 63, no. 4, pp. 1216-1229, April 2015.

\bibitem{r2x}
Y. Hama and H. Ochiai, ``A low-complexity matched filter detector with parallel interference cancellation for massive MIMO systems'' \emph{IEEE Int. Conf. Wireless Mobile Comput. Netw. Commun. (WiMob)}, Oct. 2016.

 \bibitem{r2xy}
R. Mosayebi, M. M. Mojahedian and A. Lozano,
``Linear Interference Cancellation for the Cell-Free C-RAN Uplink'', \emph{ IEEE Trans. Wireless Commun.}, vol. 20, no. 3, pp. 1544-1556, March 2021.

\bibitem{jed}
H. Song, T. Goldstein, X. You, C. Zhang, O. Tirkkonen, and C. Studer,
“Joint channel estimation and data detection in cell-free massive mu-
mimo systems,” IEEE Trans. on Wireless Commun., vol. 21, no. 6, pp.
4068–4084, 2022.

\bibitem{spa_r}
R. C. De Lamare and R. Sampaio-Neto, "Minimum Mean-Squared Error Iterative Successive Parallel Arbitrated Decision Feedback Detectors for DS-CDMA Systems," in IEEE Transactions on Communications, vol. 56, no. 5, pp. 778-789, May 2008.

\bibitem{mfsic}
P. Li, R. C. de Lamare and R. Fa, "Multiple Feedback Successive Interference Cancellation Detection for Multiuser MIMO Systems," in IEEE Transactions on Wireless Communications, vol. 10, no. 8, pp. 2434-2439, August 2011.

\bibitem{dfcc}
P. Li and R. C. De Lamare, "Adaptive Decision-Feedback Detection With Constellation Constraints for MIMO Systems," in IEEE Transactions on Vehicular Technology, vol. 61, no. 2, pp. 853-859, Feb. 2012.

\bibitem{mbdf}
R. C. de Lamare, "Adaptive and Iterative Multi-Branch MMSE Decision Feedback Detection Algorithms for Multi-Antenna Systems," in IEEE Trans. on Wir. Commun., vol. 12, no. 10, pp. 5294-5308, October 2013.

\bibitem{did}
P. Li and R. C. de Lamare, "Distributed Iterative Detection With Reduced Message Passing for Networked MIMO Cellular Systems," in IEEE Transactions on Vehicular Technology, vol. 63, no. 6, pp. 2947-2954, July 2014.

\bibitem{itic}
R. C. De Lamare, R. Sampaio-Neto and A. Hjorungnes, "Joint iterative interference cancellation and parameter estimation for cdma systems," in IEEE Communications Letters, vol. 11, no. 12, pp. 916-918, December 2007.

\bibitem{lrcc}
H. Ruan and R. C. de Lamare, "Distributed Robust Beamforming Based on Low-Rank and Cross-Correlation Techniques: Design and Analysis," in IEEE Transactions on Signal Processing, vol. 67, no. 24, pp. 6411-6423, 15 Dec.15, 2019.

\bibitem{aaidd}
R. B. Di Renna and R. C. de Lamare, "Adaptive Activity-Aware Iterative Detection for Massive Machine-Type Communications," in IEEE Wireless Communications Letters, vol. 8, no. 6, pp. 1631-1634, Dec. 2019.

\bibitem{listmtc}
R. B. Di Renna and R. C. de Lamare, "Iterative List Detection and Decoding for Massive Machine-Type Communications," in IEEE Transactions on Communications, vol. 68, no. 10, pp. 6276-6288, Oct. 2020.

\bibitem{detmtc}
R. B. Di Renna, C. Bockelmann, R. C. de Lamare and A. Dekorsy, "Detection Techniques for Massive Machine-Type Communications: Challenges and Solutions," in IEEE Access, vol. 8, pp. 180928-180954, 2020.

\bibitem{msgamp}
R. B. D. Renna and R. C. de Lamare, "Dynamic Message Scheduling Based on Activity-Aware Residual Belief Propagation for Asynchronous mMTC," in IEEE Wireless Communications Letters, vol. 10, no. 6, pp. 1290-1294, June 2021.

\bibitem{msgamp2}
R. B. Di Renna and R. C. de Lamare, "Joint Channel Estimation, Activity Detection and Data Decoding Based on Dynamic Message-Scheduling Strategies for mMTC," in IEEE Transactions on Communications, vol. 70, no. 4, pp. 2464-2479, April 2022.

\bibitem{dynovs}
Z. Shao, L. T. N. Landau and R. C. de Lamare, "Dynamic Oversampling for 1-Bit ADCs in Large-Scale Multiple-Antenna Systems," in IEEE Transactions on Communications, vol. 69, no. 5, pp. 3423-3435, May 2021.

\bibitem{llraps}
R. B. D. Renna and R. C. de Lamare, "Iterative Detection and Decoding With Log-Likelihood Ratio Based Access Point Selection for Cell-Free MIMO Systems," in IEEE Transactions on Vehicular Technology, vol. 73, no. 5, pp. 7418-7423, May 2024.

\bibitem{refidd}
T. Ssettumba, Z. Shao, L. T. N. Landau, M. Facina, P. Branco da Silva and R. C. de Lamare, "Centralized and Decentralized IDD Schemes for Cell-Free Massive MIMO Systems: AP Selection and LLR Refinement," in IEEE Access, vol. 12, pp. 62392-62406, 2024.

\bibitem{jidf}
R. C. de Lamare and R. Sampaio-Neto, "Adaptive Reduced-Rank Processing Based on Joint and Iterative Interpolation, Decimation, and Filtering," in IEEE Transactions on Signal Processing, vol. 57, no. 7, pp. 2503-2514, July 2009.

\bibitem{jiols}
R. C. de Lamare and R. Sampaio-Neto, "Reduced-Rank Space–Time Adaptive Interference Suppression With Joint Iterative Least Squares Algorithms for Spread-Spectrum Systems," in IEEE Transactions on Vehicular Technology, vol. 59, no. 3, pp. 1217-1228, March 2010.

\bibitem{jiomimo}
R. C. de Lamare and R. Sampaio-Neto, "Adaptive Reduced-Rank Equalization Algorithms Based on Alternating Optimization Design Techniques for MIMO Systems," in IEEE Transactions on Vehicular Technology, vol. 60, no. 6, pp. 2482-2494, July 2011.

\bibitem{pegbf}
A. G. D. Uchoa, C. Healy, R. C. de Lamare and R. D. Souza, "Design of LDPC Codes Based on Progressive Edge Growth Techniques for Block Fading Channels," in IEEE Communications Letters, vol. 15, no. 11, pp. 1221-1223, November 2011.

\bibitem{vfap}
J. Liu and R. C. de Lamare, "Low-Latency Reweighted Belief Propagation Decoding for LDPC Codes," in IEEE Communications Letters, vol. 16, no. 10, pp. 1660-1663, October 2012.

\bibitem{memd}
C. T. Healy and R. C. de Lamare, "Design of LDPC Codes Based on Multipath EMD Strategies for Progressive Edge Growth," in IEEE Transactions on Communications, vol. 64, no. 8, pp. 3208-3219, Aug. 2016.


\end{thebibliography}
\end{document}